\documentclass[12pt]{article}
\usepackage{amsmath}
\usepackage{graphicx}
\usepackage{enumerate}
\usepackage{natbib}
\usepackage{url}

\usepackage{setspace,lscape,longtable}
\usepackage{epsfig,graphicx}
\usepackage{mathrsfs,amsmath,amsthm,amssymb,color, verbatim}
\usepackage{multirow,makecell} 

\usepackage[T1]{fontenc}
\usepackage[utf8]{inputenc}
\usepackage{authblk}

\usepackage{geometry}
\usepackage{float}
\usepackage{array}
\usepackage{lscape}
\usepackage{bm}

\usepackage{comment}
\usepackage{setspace}
\usepackage{amsfonts}
\usepackage{indentfirst}
\usepackage{booktabs}
\usepackage{caption}
\usepackage{graphicx, subfig}
\usepackage[linesnumbered,ruled,vlined]{algorithm2e}
\SetKwInput{KwInput}{Input}                
\SetKwInput{KwOutput}{Output}

\newtheorem{theorem}{Theorem}
\newtheorem{lemma}{Lemma}
\newtheorem{remark}{Remark}
\newtheorem{proposition}{Proposition}

\def\1{\mathbf{1}}

\def\mD{\mathcal D}

\def\mR{\mathbb{R}}

\def\hat{\widehat}

\def \tbe_t{\tilde{\bm{\epsilon}}_t}

\def\beq{\begin{equation}}
\def\eeq{\end{equation}}
\def\ben{\begin{equation*}}
\def\een{\end{equation*}}
\def\bea{\begin{eqnarray}}
\def\eea{\end{eqnarray}}

\def\bda{\begin{eqnarray*}}
\def\eda{\end{eqnarray*}}

\def\bet{\begin{theorem}}
\def\eet{\end{theorem}}
\def\bel{\begin{lemma}}
\def\eel{\end{lemma}}
\def\bep{\begin{proposition}}
\def\eep{\end{proposition}}
\def\bg{\begin{figure}[tbph]\begin{center}}
\def\eg{\end{center}\end{figure}}

\def\bc{\begin{center}}
\def\ec{\end{center}}

\newcommand{\blind}{1}

\numberwithin{equation}{section}

\begin{document}

\def\spacingset#1{\renewcommand{\baselinestretch}%
{#1}\small\normalsize} \spacingset{1}


\if1\blind
{
  \title{\bf A Goodness-of-Fit Test for Sparse Networks}

  \author[1]{Yujia Wu}
  \author[1]{Wei Lan\thanks{Corresponding author Wei Lan, Email: lanwei@swufe.edu.cn.\hspace{.2cm}}$^{,}$}
  \author[2]{Long Feng}
  \author[3]{Chih-Ling Tsai}

  \affil[1]{School of Statistics and Data Science and Center of Statistical Research, Southwestern University of Finance and Economics, Chengdu, China}
  \affil[2]{School of Statistics and Data Science, Nankai University, Tianjin, China}
  \affil[3]{Graduate School of Management, University of California, Davis, CA}
  \renewcommand*{\Authands}{ and }  
  \date{}  
  \maketitle
} \fi

\if0\blind
{
  \bigskip
  \bigskip
  \bigskip
  \begin{center}
    {\LARGE\bf A Goodness-of-Fit Test for Sparse Networks}
\end{center}
  \medskip
} \fi

\bigskip
\begin{abstract}
The stochastic block model (SBM) has been widely used to analyze network data.
Various goodness-of-fit tests have been proposed to assess the adequacy of model structures.
To the best of our knowledge, however, none of the existing approaches are applicable for
sparse networks in which the connection probability of any two communities
is of order $n^{-1}\log n$, and the number of communities is
divergent. To fill this gap, we propose a novel goodness-of-fit test for the stochastic block model.
The key idea is to construct statistics by sampling the maximum entry-deviations of the adjacency matrix that the negative impacts of network sparsity are alleviated by the sampling process.
We demonstrate theoretically that the proposed test statistic converges to the Type-I extreme value distribution under the null hypothesis
regardless of the network structure. Accordingly, it can be applied to both dense and sparse networks.
In addition, we obtain the
asymptotic power against  alternatives.
Moreover, we  introduce
a bootstrap-corrected test statistic to
improve the finite sample performance, recommend
an augmented test statistic to increase the power, and extend the proposed test to the degree-corrected SBM.
Simulation studies and two empirical examples with both dense and sparse networks
indicate that the proposed method performs well.

\end{abstract}

\vfill

\spacingset{1.9} 
\section{Introduction}
\label{sec:intro}

Due to the rapid development and application of information technology,
immense troves of network data have been created
across various fields, such as finance and business (\citealt{katona2011network}; \citealt{di2019relevance}; \citealt{hardle2016tenet}), social science (\citealt{bramoulle2009identification}; \citealt{jin2024mixed}; \citealt{andrikopoulos2016four})
engineering (\citealt{agarwal2013traffic}), machine learning (\citealt{musumeci2018overview}).
To make full use of network data, it is critical to understand their generation mechanisms.
To this end, the stochastic block model (SBM) has been widely studied and used since the pioneering work of \cite{holland1983stochastic}; see, e.g., \cite{snijders1997estimation}, \cite{nowicki2001estimation}, \cite{bickel2009nonparametric}, \cite{rohe2011spectral}, \cite{choi2012stochastic}, \cite{jin2015fast}, \cite{zhang2016minimax}, \cite{hu2021using} and \cite{jochmans2024nonparametric}. 
A good review paper on SBM can be found in \cite{lee2019review}.

The SBM basically assumes that the network nodes can be decomposed into
$K$ non-overlapping communities, where  each node belongs to one specific community.
In addition, the edge connecting any two nodes depends only on the communities in which they are located.
Consider an undirected network containing $n$ nodes existing within $K$ communities. For each node $i\in\{1, \cdots, n\}$, $g(i)\in\{1, \cdots, K\}$ represents the community to which node $i$ belongs.
Let $A=(a_{ij})\in\mR ^{n\times n}$ be a symmetric adjacency matrix, where $a_{ij}=1$ if two nodes $i$ and $j$ are connected and
$a_{ij}=0$ otherwise. For the sake of  completeness, the diagonal elements of $A$ are set to zero.
We then assume that each $a_{ij}$, for $i>j$, is independently generated from
the Bernoulli distribution with the probability $Q_{g(i)g(j)}$, where $0\le Q_{g(i)g(j)}\le 1$ is the connection probability of nodes $i$ and $j$.

When the community number $K$ is known,
various approaches have been proposed to estimate the community membership vector $g=(g(1),\cdots, g(n))^\top$;
see, e.g., the modularity-based method (\citealt{newman2006modularity}),
the profile-likelihood maximization method (\citealt{bickel2009nonparametric}), the pseudo-likelihood maximization method (\citealt{amini2013pseudo}),
the variational method (\citealt{daudin2008mixture}), and the spectral
clustering method (\citealt{rohe2011spectral} and \citealt{jin2015fast}).
In addition, some researchers have established the
theoretical properties of the estimator of $g$ under the condition $n\min_{i,j}Q_{g(i)g(j)}\to\infty$ or even under stronger conditions; see, e.g., \cite{choi2012stochastic}, \cite{rohe2011spectral}, \cite{zhao2012consistency}, \cite{sarkar2015role}, \cite{jin2015fast}, \cite{lei2015consistency}, and \cite{zhang2016minimax}.

To broaden the usefulness of the SBM,
it is crucial to check the model adequacy. Hence,
\cite{bickel2016hypothesis} used the largest eigenvalue of the centered and scaled adjacency matrix to test the Erd\H{o}s-R\'{e}nyi model, that is, the SBM with
one community,
\cite{karwa2016exact} developed a finite-sample Monte Carlo method to assess the SBM.
It is worth noting that the aforementioned testing methods
rely on two
stringent conditions: (i)
the number of communities $K$ is finite, and (ii) the network is dense, such that
$n\min_{i,j} Q_{g(i)g(j)}/\log n\rightarrow \infty$.

Relaxing Condition (i) to accommodate a divergent $K$,
\cite{lei2016goodness} proposed a goodness-of-fit test of the SBM. Specifically, he employed the largest singular value of the centered and rescaled adjacency matrix $A$, and  then
obtained the asymptotic null distribution under the condition $K= o(n^{1/6})$.
To further relax this condition, \cite{hu2021using} developed a test statistic based on the maximum entry-wise deviations of the centered and rescaled adjacency matrix $A$,
which allows the number of communities to grow with $K=o(n/\log^2n)$.
Although the methods of \cite{lei2016goodness} and \cite{hu2021using} are applicable for a divergent $K$,
these two methods require the condition $\min_{i, j} Q_{g(i)g(j)}\geq c$ for a constant $0<c<1$ and
it is expected that these two methods are  inapplicable when Condition (ii) is not valid.

It is worth noting that sparse networks often appear in practical applications;
see, e.g., \cite{amini2013pseudo}; \cite{sarkar2015role}; \cite{wang2017likelihood}; and \cite{jing2022community}.
To get more insight regarding sparsity in networks, we consider two datasets given below.
(A) The Sina Weibo network consists of 2,580 nodes, which was studied by \cite{wu2022inward}.
In this network, $a_{ij}=1$ if user $i$ follows user $j$. The network density
is 1.3\%, calculated by the formula $\sum_{i,j} a_{ij}/(n(n-1))$.
(B) The co-authorship network contains 3,607 nodes, which was analyzed by \cite{ji2016coauthorship}. The connection between any two nodes is defined as follows: $a_{ij}= 1$ if and only if authors $i$ and $j$ have at least two coauthored papers. The network density is approximately 0.012\%.
To assess the sparsity of the above two networks, let $D_n$ denote the density of the network. We then calculate the ratio $n D_n/\log n$,  resulting in a value of 4.273 for dataset (A) and 0.053 for dataset (B).
Since both values of $nD_n/\log n$ are not large,
 these two datasets are expected to satisfy $D_n = O(n^{-1}\log n)$ and can be considered sparse.

In general, one can employ the SBM to analyze the above sparse datasets.
Some researchers have developed test statistics and obtained their asymptotic null distributions in sparse networks.
For example, \cite{han2023universal} proposed a residual subsampling (RIRS) approach for testing and estimating $K$ with the network density being $O(n^{-1+\epsilon})$
for some arbitrarily small $\epsilon>0$. \cite{zhang2020adjusted} developed an adjusted chi-square statistic, which is applicable for testing SBM with an average degree no smaller than
$O(\log n)$. However, the aforementioned two tests are only applicable for fixed $K$.
\cite{han2023universal} emphasized in their Section 2.1 that they required the mean of the adjacency matrix $A$ to be of fixed rank, which implicitly implies that
$K$ is fixed. In addition,
as discussed in Section 2.2 of \cite{zhang2020adjusted}, they also required a constant value of $K$ in order to ensure the convergence
of their adjusted chi-square test statistic to the standard normal distribution. Our simulation results in Section 3 confirm their limitation;
both methods exhibit size distortions for large $K$.
To the best of our knowledge, there is no goodness-of-fit
test for evaluating the suitability of the SBM for a sparse network with a
divergent $K$.

The aim of this paper is to develop a new test statistic that allows for testing the number of communities $K$ and the mapping vector $g=(g(1),\cdots, g(n))^\top$,
without regard to the sparsity of the network or the divergence of $K$.
To obtain this test, we
not only allow $K$ to be  divergent, but also simply require the condition,
$n\min_{i, j} Q_{g(i)g(j)}\rightarrow \infty$.
This is a relaxed version of the Condition (ii) mentioned earlier.
The procedures of the proposed method are as follows. We first
adopt the approach of \cite{hu2021using} to calculate the entry-wise
deviation $\hat\rho_{iv}$ for each node $i$ in community $v\in\{1,\cdots, K\}$, which is constructed based on the average of the standardized summation of $a_{ij}$ for all nodes $j\not=i$ within the community $v$.
Subsequently, for any fixed community  $v\in\{1,\cdots, K\}$, we randomly draw $B$ samples from $n$ nodes with replacement and then calculate $\hat\psi_{\cdot v}=\sum_{i=1}^B\hat \rho_{i v}/\sqrt{B}$.
Repeating this sampling procedure
 $M$ times for each of the $K$ communities,
 we  obtain the $M$ values of $\hat\psi_{\cdot v}$, and
denote them $\hat\psi_{mv}$ for $m=1,\cdots, M$. Finally, the new test statistic,
$\max_{1\le m\le M,1\le v\le K}|\hat\psi_{mv}|$, is constructed.
By the central limit theorem, we can  demonstrate that $\hat\psi_{mv}$ converges to the standard normal distribution $N(0,1)$ plus  an additional term of the order $O_p(1/\sqrt{B\log n})$.
By setting $B$ to be divergent, the additional term with the order $O_p(1/\sqrt{B\log n})$ becomes negligible.
As a result, we expect that $\max_{1\le m\le M,1\le v\le K}|\hat\psi_{mv}|$ converges to an extreme value distribution under certain conditions regardless of dense networks or sparse networks.

Although implementing our proposed test is not complicated,
establishing its theoretical properties is challenging. This is because
the $\hat\psi_{mv}$'s are correlated due to the replacement samples from $\hat\rho_{iv}$.
As a result,
 $\max_{m, v}|\hat\psi_{mv}|$ is no longer the maximum of a sequence of independent variables, and its asymptotic distribution
 is more difficult to demonstrate than that of \cite{hu2021using}.
In the paper, we overcome challenges
and show
 that $\max_{1\le m\le M,1\le v\le K}|\hat\psi_{mv}|$ converges to the Type-I extreme value distribution.
In addition, according to the aforementioned explanations, our approach comprises \cite{hu2021using}'s method and the sampling with
replacement procedure. Accordingly, our proposed method is versatile for both dense and sparse networks. In contrast, \cite{hu2021using}'s
method is tailored for dense networks. Consequently, our proposed test exhibits greater
robustness in practical application.
Since the test statistic's rate of convergence to the extreme distribution may be slow, we further consider a finite-sample bootstrap correction by generating bootstrap samples to improve the finite sample performance.
Finally, we introduce
an augmented test statistic to increase the power and extend the proposed test to the degree-corrected SBM.

The remainder of this paper is organized as follows. Section 2 introduces the proposed test and studies asymptotic null distribution, the power properties,  the bootstrap-corrected method, and an augmented test statistic.
The simulation studies are presented in Section 3,
and they indicate that the proposed method outperforms those in \cite{hu2021using}, \cite{lei2016goodness}, \cite{han2023universal} and \cite{zhang2020adjusted} under sparse networks.
Furthermore,
our augmented bootstrap-corrected test is comparable or superior to those methods under both dense and sparse networks.
Section 4 provides two real data examples with the dense and sparse networks, separately. As found in simulation studies, our method works well empirically regardless the network density.
Section 5 discusses the degree-corrected SBM and provides concluding remarks. All the technical details and additional simulation results are relegated to the supplementary material.

\section{Model and theoretical properties}
\label{sec:meth}

\subsection{Hypothesis testing}

Consider a network of $n$ nodes with the adjacency matrix $A$.
Our focus is to assess whether this network  can be successfully fitted
by a stochastic block model with $K_0$ communities and
a membership vector $g_0$. Let
$K$ and $g$ denote the  true number of communities and the true membership vectors, respectively.
Based on the discussions in \cite{hu2021using},
the goodness-of-fit test for the SBM is equivalent to testing the following two hypotheses:
\begin{itemize}
\item [(I)] $H_{0k}: K=K_0$ v.s. $H_{1k}: K>K_0$ {\mbox{~and~}}  (II) $H_{0g}:g=g_0$ v.s. $H_{1g}:g\not=g_0$.
\end{itemize}
It is worth noting that, for hypothesis test (I), we consider only a one-sided alternative $K>K_0$.
The main reason is that each SBM with $K$ communities can be split into $K_0>K$ communities.
Accordingly, we consider a one-sided alternative with nodes being partitioned into less than $K$ communities in hypothesis test (I) to
guarantee the power of goodness-of-fit tests (e.g., \citealt{lei2016goodness}; \citealt{chen2018network}; \citealt{wang2017likelihood}).

To test the null hypotheses in (I) and (II), we employ the maximum likelihood estimation method
to estimate the probability matrix $Q=(Q_{g(i)g(j)})$ under the null hypotheses.
Consider
the probability mass function of the adjacency matrix $A$,
$P_{g,Q}(A)=\prod_{1\leq i<j\leq n}Q^{a_{ij}}_{g(i)g(j)}(1-Q^{a_{ij}}_{g(i)g(j)})^{1-a_{ij}}.$
Denote $n_{uv}=\sum_{i=1}^n\sum_{j\not=i}\mathbf{1}\{g(i)=u,g(j)=v\}$ and $m_{uv}=\sum_{i=1}^n\sum_{j\not=i}a_{ij}\mathbf{1}\{g(i)=u,g(j)=v\}$.
Then the corresponding log-likelihood function is
$\ell(A|Q,g)=\frac{1}{2}\sum_{u,v=1}^K\big\{m_{uv}\text{log}~Q_{uv}+(n_{uv}-m_{uv})\text{log}(1-Q_{uv})\big\}.$
Given the
hypothetical number of communities $K_0$ and the hypothetical membership vector $g_0$,
we can obtain the maximum likelihood estimator of $Q$ by maximizing $\ell(A|Q,g_0)$. It is
\beq\label{eq:hatB}
\hat Q_{uv}^{g_0}=
\left\{\begin{array}{l}
\frac{\sum_{i\in g_0^{-1}(u),j\in g_0^{-1}(v)}a_{ij}}{|g_0^{-1}(u)||g_0^{-1}(v)|},\quad 1\leq u\not=v\leq K_0,\\
\frac{\sum_{i\not=j\in g_0^{-1}(u)}a_{ij}}{|g_0^{-1}(u)|(|g_0^{-1}(u)|-1)},\quad u=v,
\end{array}\right.
\eeq
where $g_0^{-1}(u)=\{i:1\le i\le n,g_0(i)=u\}$ and $|g^{-1}_0(u)|$ is the number of nodes within the community $u$.
Note that, to test the null hypothesis in (I),  we replace $g_0$ in (\ref{eq:hatB}) with its consistent estimator $\hat g$;
see, e.g., the estimator obtained using \cite{lei2015consistency}'s spectral clustering method.
We next adopt \cite{hu2021using}'s approach and introduce the maximum entry-wise deviation.

\subsection{Maximum entry-wise deviation}

To test the null hypotheses in (I) and (II)  with a divergent $K$, \cite{hu2021using} proposed the entry-wise deviation,
\beq
\hat\rho_{iv}=\frac{1}{\sqrt{|g_0^{-1}(v)/\{i\}|}}\sum_{j\in g_0^{-1}(v)/\{i\}}\frac{a_{ij}-\hat Q_{g_0(i)g_0(j)}^{g_0}}{\sqrt{\hat Q_{g_0(i)g_0(j)}^{g_0}(1-\hat Q_{g_0(i)g_0(j)}^{g_0})}}
,\label{eq:Zhang.hatrho}
\eeq
where $g_0^{-1}(v)/\{i\}$ is the set of nodes contained in $g_0^{-1}(v)$
excluding node $i$.
They then obtained the maximum entry-wise deviation
$L_n(K_0,g_0)\overset{\Delta}{=}\max_{1\le i\le n,1\le v\le K_0}|\hat\rho_{iv}|$.
When the network is dense such that $\min_{i,j} Q_{g(i)g(j)}\geq c$ for a constant $0<c<1$, \cite{hu2021using} established the theoretical properties of $L_n(K_0,g_0)$.

As noted in Section 6 of \cite{hu2021using}, their proposed test is not applicable for sparse networks. For the sake of illustration,
we consider a sparse network with $c_{\min}n^{-1}\log n\le \min_{i,j} Q_{g(i)g(j)}\le \max_{i,j} Q_{g(i)g(j)}\le c_{\max}n^{-1}\log n$, where $c_{\min}$ and $c_{\max}$ are two finite positive constants.
Under this scenario, $\hat\rho_{iv}$ is a summation of $c_{\rho}\log n$ random terms based on (\ref{eq:Zhang.hatrho}), where $\log n$ is related to the order of network density (i.e., $n^{-1}\log n$),
and $c_{\rho}$ is a constant between $c_{\min}$ and $c_{\max}$.
Therefore, $\hat\rho_{iv}$ can be expressed as a standard normal distribution $N(0,1)$ plus an additional term $O_p(1/\sqrt{\log n})$.
Accordingly, it can be demonstrated that the test statistic of \cite{hu2021using}, denoted as $L_n^2(K,g)-2\log(Kn)+\log\log(Kn)$, converges to an extreme value distribution plus an additional term of $O_p(1)$.
Consequently, the test statistic proposed by \cite{hu2021using} fails to converge to the Gumbel distribution under the sparse network setting, while it is true under the dense network setting demonstrated  by \cite{hu2021using}. This finding is supported by our simulation experiments given in Subsection 3.2. This motivates us to develop a new test statistic that performs well without regard to the network density (or sparsity).

\begin{remark}
	
	It is worth noting that the entrywise deviation statistic $\hat{\rho}_{iv}$, proposed by \citet{hu2021using},  has also been extended to various network testing problems, such as
 two-sample testing; see, for example, \cite{wu2024atwosample}, \cite{wu2024twosample} and  \cite{Fu2025twosample}. These methods are tailored for dense networks \citep{Fu2025twosample, wu2024atwosample}, or for sparse regimes satisfying $ \min_{i,j} Q_{g(i)g(j)}\ge c_{\min}n^{-1}\log n$ for a constant $c_{\min}>0$ \citep{wu2024twosample}.
	However, it is important to note that these methods are not applicable to the testing problems considered in our work, as they focus on assessing whether two networks differ in their community structures. In these works, they assume that both networks are already generated from a SBM with a pre-specified number of communities $K$. Thus, these works are not capable of simultaneously testing the two hypotheses (I) and (II) considered in our paper.
	
\end{remark}

\subsection{Maximum sampling entry-wise deviation test statistic}

In this subsection, we propose a novel test statistic that is applicable
for both dense and sparse networks.
The key idea is to create the maximum entry-wise deviation by integrating the entry-wise deviation $\hat\rho_{iv}$ with the sampling process.
Specifically, we calculate  the $\hat\rho_{iv}$'s
based on \cite{hu2021using}.  For any fixed community $v\in \{1, \cdots, K_0\}$, we   randomly draw $B$ samples from $\{\hat\rho_{1v},\cdots,\hat\rho_{nv}\}$ at the $m$-th realization ($m=1,\cdots,M$), 
and denote them $\{\hat\rho_{m_1v}, \cdots, \hat\rho_{m_Bv}\}$. Subsequently,
calculate
\beq
\hat\psi_{mv}=\frac{1}{\sqrt{B}}\sum_{h=1}^B \hat\rho_{m_hv}.\label{eq:hat.gamma}
\eeq 
Repeating this procedure $M$ times for each community $v$, we obtain the
sequence of $\hat\psi_{mv}$'s for $m=1,\cdots, M$.
Under the null hypotheses in (I) and (II), one expects all $\hat\rho_{iv}$ values to be small as noted by \cite{hu2021using}.
Accordingly, $\hat\psi_{mv}$ should also be small, which motivates us to
propose the maximum sampling entry-wise deviation based on the maximum value of $\hat\psi_{mv}$'s across $m$ and $v$,
\beq\label{eq:F0}
\Gamma_n(K_0,g_0)\overset{\Delta}{=}\max_{1\le m\le M,1\le v\le K_0}|\hat\psi_{mv}|.
\eeq
By setting $B=1$ and choosing $M$ to be sufficiently large such that
all the $\hat\rho_{iv}$'s can be selected,
$\Gamma_n(K_0,g_0)$ is then identical to $L_n(K_0,g_0)$.

It is worth noting that the above
sampling process can alleviate the negative impact of network sparsity
by letting $B$ be divergent in $n$.
Considering a
sparse network with $c_{\min}n^{-1}\log n\le \min_{i,j} Q_{g(i)g(j)}\le \max_{i,j} Q_{g(i)g(j)}\le c_{\max}n^{-1}\log n$ as discussed in Section 2.2,
 $\hat\rho_{iv}$ is a summation of $c_{\rho}\log n$ random terms.
Accordingly, (\ref{eq:hat.gamma})  can be expressed as a summation of infinite $c_{\rho} B\log n$ random terms as $B\to\infty$ along with $n$.
By the central limit theorem, $\hat\psi_{mv}$ can be expressed as the standard normal distribution $N(0,1)$ plus a term
$O_p(1/\sqrt{B\log n})$. By  Lemma 1 given in the supplementary material, we can show that
$\Gamma_n(K_0,g_0)=O_p\big\{\sqrt{\log(MK_0)}\big\}\big\{O_p(1)+O_p(1/\sqrt{B\log n})\big\}=O_p\big\{\sqrt{\log(MK_0)}\big\}+o_p(1)$ under Conditions (C2) and (C3)
below.
Thus, the term $O_p(1/\sqrt{B\log n})$ is negligible in deriving the asymptotic distribution of $\Gamma_n(K_0,g_0)$.
As a result, we expect that
$\Gamma_n(K_0,g_0)$ follows an extreme value distribution under certain conditions,
and it is applicable to both dense and sparse networks.
To rigorously derive the theoretical properties of $\Gamma_n(K_0,g_0)$, we introduce the following technical conditions.

\begin{itemize}
\item [(C1)] Assume that $Q$ has no identical rows. In addition, assume $Q_{g(i)g(j)}\in(0,1)$ for $1\le i,j\le n$ uniformly, and $n\min_{i,j}Q_{g(i)g(j)}\to\infty$ as $n\rightarrow \infty$.

\item [(C2)]  Assume that there exist two finite positive constants $\xi_1$ and $\xi_2$ such that
$$\xi_1n/K<\min_{1\le u\le K}|g^{-1}(u)|\le\max_{1\le u\le K}|g^{-1}(u)|<\xi_2n^2/(K^2\log^2n),$$
uniformly for any $n$, where $|g^{-1}(u)|$ is the number of nodes contained in the $u$-th community defined in (\ref{eq:hatB}).

\item [(C3)]
Assume that
(i) $B=\max[c_1q^{-1/2}, c_2(n/\log n)^{\xi_3-\xi_c})]$ for some finite positive constants $c_1$, $c_2$ and $\xi_c$,
(ii) $K=o\{(n/\log n)^{\xi_3}\}$, and (iii) $MK=O(n^{\xi_4})$, where $q=\max_{i,j} Q_{g(i)g(j)}$, and $0<\xi_c\le\xi_3<1/2$ and
$0<\xi_4<1$ are finite constants.
In addition, assume that $\min\{n, M, B\}\rightarrow \infty$.
\end{itemize}

Condition (C1) requires that the probability matrix $Q$ is identifiable with entries all bounded in $(0,1)$.
This condition  is similar to, but weaker than, that of \cite{lei2016goodness}, \cite{wang2017likelihood}, and \cite{hu2021using}.
Note that if all $Q_{g(i)g(j)}$s have the same order, then
$Q_{g(i)g(j)}$ shares the same order with the network density $D_n$.
Accordingly, this condition immediately implies that the network density $D_n$ satisfies $n D_n\rightarrow \infty$ and allows the network density to approach zero more slowly
than $n^{-1}$ as $n\to\infty$. In addition, if $\hat Q^g_{g(i)g(j)}=0$, then the term $\big(a_{ij}-\hat Q_{g_0(i)g_0(j)}^{g_0}\big)/\big(\sqrt{\hat Q_{g_0(i)g_0(j)}^{g_0}(1-\hat Q_{g_0(i)g_0(j)}^{g_0})}\big)$ in the summation of (\ref{eq:Zhang.hatrho}) is removed, and our analysis remains valid.
For the sake of simplicity, we omit the case which $Q_{g(i)g(j)}=0$ in this paper.
Condition (C2) is a mild condition that requires
the size of the smallest community to be at least proportional to $n/K$. Furthermore, Condition (C2) imposes an upper bound on the largest community size, which was also considered by \cite{zhang2016minimax}, \cite{gao2018community}, and \cite{hu2021using}.
This condition is used to control the estimation bias between $\hat Q^g_{g(i)g(j)}$ and its population version $Q_{g(i)g(j)}$,
$\max_{1\le i\le n,1\le v\le K}\left|\frac{1}{\sqrt{|g^{-1}(v)/\{i\}|}}\sum_{j\in g^{-1}(v)/\{i\}}\frac{Q_{g(i)g(j)}-\hat Q^g_{g(i)g(j)}}{\sqrt{Q_{g(i)g(j)}(1-Q_{g(i)g(j)})}}\right|.$
Specifically, the condition ensures that this estimation bias converges to zero in probability.
Condition (C3) identifies the orders of $K$, $B$, and $M$. Intuitively, $B$ and $M$ cannot be too large; otherwise, the correlation between $\hat\psi_{mv}$'s  may be large. Note that Condition (C3)-(ii) allows  $K$  to grow more slowly along $n$ than the assumption that  $K=o{(n/\log^2 n)}$, which was  proposed by \cite{hu2021using}. In addition, \cite{han2023universal} required $K$ to be a fixed value, although they introduced less stringent density conditions than those in \cite{lei2016goodness}, \cite{wang2017likelihood}, and \cite{hu2021using}. Consequently, in comparison to \cite{lei2016goodness}, \cite{wang2017likelihood}, \cite{hu2021using}, and \cite{han2023universal}, our method not only exhibits the adaptability to a broad spectrum of network densities, but also allows $K$ to be divergent.
Based on the above three conditions,
we next obtain the theoretical properties of  
$\Theta_{n}=\Gamma_n^2(K_0,g_0)-2\log(MK_0)+\log\log(MK_0).$ 

\bet\label{them-size}
Assume Conditions (C1)-(C3) hold. Under the null hypotheses of $K=K_0$ and $g=g_0$,
we have that,  for any $x\in\mR$,
\beq
\lim_{n\to\infty}P\Big\{\Theta_n\le x\Big\}
=\exp\big(-\frac{1}{\sqrt{\pi}}e^{-x/2}\big), \label{eq:size}
\eeq
where the right hand side of (\ref{eq:size}) is the CDF of
the Type-I extreme value distribution  (i.e., Gumbel distribution) with $\mu = -2\log(\sqrt{\pi})$ and $\beta=2$.
\eet

Applying Theorem~\ref{them-size}, we can test the null hypotheses in (I) and (II).
To test the null hypothesis in (I), we replace $g_0$ with its consistent estimator $\hat g$.
This estimator can be obtained regardless the
network being dense, $n\min_{i, j} Q_{g(i)g(j)}/\log n\rightarrow \infty$, (see, e.g., \citealt{daudin2008mixture}; \citealt{lei2015consistency}; \citealt{lei2017generic}; \citealt{gao2017achieving} and \citealt{gao2018community}) or
sparse, $n\max_{i, j} Q_{g(i)g(j)}=O(\log n)$, (see, e.g., \citealt{zhang2016minimax}; \citealt{jing2022community}).
The resulting test statistic is
$\Theta_{n,1}=\Gamma_n^2(K_0,\hat g)-2\log(MK_0)+\log\log(MK_0).$
We can demonstrate theoretically that $\Theta_{n,1}$ has the same asymptotic null distribution as $\Theta_n$ under proper conditions; see Corollary S.1 and its proof in Section S.7
of the supplementary material.
Consequently, we reject the null hypothesis, $H_{0k}:K=K_0$, if $\Theta_{n,1}>q_{\alpha}$, where $q_{\alpha}=-2\log\{-\sqrt{\pi}\log(1-\alpha)\}$
is the $(1-\alpha)$-th quantile of the Type-I extreme distribution with the distribution function $\exp\big(-e^{-x/2}/\sqrt{\pi}\big)$.

As for testing the null hypothesis in (II),  we can directly calculate the test statistic:
$\Theta_{n,2}=\Gamma_n^2(K_0,g_0)-2\log(MK_0)+\log\log(MK_0).$
This is because $g_0$ can yield its corresponding $K_0$.
Thus, we can reject the null hypothesis $H_{0g}:g=g_0$ when $\Theta_{n,2}>q_{\alpha}$ at the given significance level $\alpha$.

For ease of understanding and practical applications, we present Algorithm 1 below with five steps to calculate the test statistic $\Theta_{n,2}$ for testing (II).
To obtain $\Theta_{n,1}$ for testing (I), one can replace $g_0$ in
$\Theta_{n,2}$ with its consistent estimator $\hat g$ calculated  under $K=K_0$.

\begin{algorithm}[htbp]
  \KwInput{Adjacency matrix $A$, the mapping vector $g_0$,
the number of communities $K_0$ produced by its corresponding $g_0$, the number of realizations $M$, the number of random samples $B$, and the significance level $\alpha$.}
  \KwOutput{Test statistic $\Theta_{n,2}$ and the decision in regard to the  null hypothesis $H_{0g}$ being rejected or not.}

  Calculate the probability matrix $\hat Q$ using  (\ref{eq:hatB}).

  Calculate $\hat\rho_{iv}$ using (\ref{eq:Zhang.hatrho}) for each $1\le i\le n$ and $1\le v\le K_0$.

 At the  $m$-th realization for $m=1,\cdots M$, randomly draw $B$ samples from $\{\hat\rho_{1v},\cdots,\hat\rho_{nv}\}$ obtained in Step 2 for each $1\le v\le K_0$, and calculate $\hat\psi_{mv}$ using (\ref{eq:hat.gamma}).

 Find the maximum value of $\hat\psi_{mv}$ for $1\le m\le M$ and $1\le v\le K_0$, and obtain $\Gamma_n(K_0,g_0)$ using (\ref{eq:F0}).

  Calculate $\Theta_{n,2}=\Gamma_n^2(K_0,g_0)-2\log(MK_0)+\log\log(MK_0)$ and reject the null hypothesis $H_{0g}:g=g_0$ if $\Theta_{n,2}>q_{\alpha}$.
\caption{Maximum Sampling Entry-Wise Deviation Test Statistic}
\end{algorithm}

\begin{remark}
Although the network considered in this paper requires $n\min_{i,j}Q_{g(i)g(j)}\to\infty$, it can be shown that Theorem 1 is also applicable even under
the condition of extreme network sparsity,  $n\max_{i,j}Q_{g(i)g(j)}<\infty$, as long as a consistent estimator $\hat g$ can be obtained.
For example, a consistent estimator $\hat g$ can be obtained for extremely sparse networks when $K=2$
(see, e.g., \citealt{mossel2015reconstruction}; \citealt{mossel2018proof}; \citealt{chin2015stochastic}).
\end{remark}

\begin{remark}
In addition to the testing methods mentioned above, there are selection methods that can be used to choose $K$.
\cite{wang2017likelihood} proposed a likelihood-based approach to select the optimal SBM, while \cite{li2020network} developed a model-free edge cross-validation procedure
for selecting $K$ under SBM. However, these two methods require the stringent conditions: (i) the number of communities $K$ is finite, and (ii) the network is dense with $n\min_{i,j} Q_{g(i)g(j)}/\log n\rightarrow \infty$. To relax condition (ii), \cite{chen2018network} proposed a network cross-validation (NCV)
approach to select $K$, but their method still requires condition (i). In contrast, \cite{chatterjee2015matrix} introduced an universal singular
value thresholding (USVT) estimation method that allows  $K$ to diverge.
However, his method is not suitable for sparse networks.
It is noteworthy that selection- and estimation-type  methods are
 not applicable for testing the model adequacy.
Thus, we do not compare our testing method with the aforementioned four methods in simulation and real data analyses.
\end{remark}

\subsection{The asymptotic power}

This subsection investigates the asymptotic power of the proposed test statistic under a class of alternatives
modified from \cite{hu2021using}.
For a given hypothetical membership vector $g_0$, we
 define the probability matrix $Q^{g_0}$ as
\beq
Q_{uv}^{g_0}=
\left\{\begin{array}{l}
\frac{\sum_{i\in g_0^{-1}(u),j\in g_0^{-1}(v)}Q_{g(i)g(j)}}{|g_0^{-1}(u)||g_0^{-1}(v)|},\quad 1\leq u\not=v\leq K_0,\\
\frac{\sum_{i\not=j\in g_0^{-1}(u)}Q_{g(i)g(j)}}{|g_0^{-1}(u)|(|g_0^{-1}(u)|-1)},\quad u=v.
\end{array}\right.\label{eq:Qg0}
\eeq
The differences between the null (i.e., $K=K_0$ and $g=g_0$) and alternative (i.e., $K>K_0$ and $g\not=g_0$) hypotheses can be characterized by the differences between
$Q=Q^g$ and $Q^{g_0}$.
To derive the asymptotic power under the alternative hypotheses, \cite{hu2021using} introduced an
alternative subset defined as $\{(K_0,g_0):K_0<K,\ell^{*}(K_0,g_0)/\sqrt{\log n}\to\infty\}$,
where $\ell^{*}(K_0,g_0)=\max_{1\le i\le n,1\le v\le K_0}\Big|\sum_{j\in g_0^{-1}(v)}\big(Q_{g(i)g(j)}-Q^{g_0}_{g_0(i)g_0(j)}\big)/\sqrt{|g_0^{-1}(v)|}\Big|$.

However, $\ell^{*}(K_0,g_0)$ is unable to capture the differences between $Q$ and $Q^{g_0}$ in  sparse networks since
both $Q$ and $Q^{g_0}$ can
decay to zero.
For the sake of illustration, we consider sparse networks with $\max_{i,j}Q_{g(i)g_(j)}=O(\log n/n)$.
Under Condition (C2*) given below, it can be shown that $\ell^{*}(K_0,g_0)=O\big\{(\max_u |g_0^{-1}(u)|)^{1/2}\log n/n\big\}=O(K_0^{-1})$, which implies that $\ell^{*}(K_0,g_0)/\sqrt{\log n}\not\to\infty$.

In order to obtain an alternative set that can accommodate sparse networks, we introduce a new measure to gauge the disparity between $Q$ and $Q^{g_0}$. Specifically, for any $i\in g^{-1}(u^*)$ with $u^*=1,\cdots,K$, we define
$q_{vu^*}=\big|\frac{1}{\sqrt{\big|g_0^{-1}(v)/\{i\}\big|}}\sum_{j\in g_0^{-1}(v)/\{i\}}\frac{Q_{g(i)g(j)}-Q^{g_0}_{g_0(i)g_0(j)}}{\sqrt{Q^{g_0}_{g_0(i)g_0(j)}(1-Q^{g_0}_{g_0(i)g_0(j)})}}\big|$,
where $v=1,\cdots, K_0$. In addition, set
\beq\label{eq:q(v)}
q_{v}\doteq \sum_{u^*=1}^K \frac{|g^{-1}(u^*)|}{n}q_{vu^*}.
\eeq
We then define the alternative set for $K_0$ and $g_0$ as follows:
\[\mD(K,g,Q)=\Big\{(K_0,g_0):K_0<K, \sqrt{B}\max_{1\le v\le K_0}q_{v}/\sqrt{\log (MK_0)}\to \infty \Big\},\]
where $B$ and $M$ are selected in accordance with Condition (C3). The $\mD(K,g,Q)$ indicates that the order of the maximum mean difference between $Q$ and $Q^{g_0}$ within the $K_0$ blocks is  greater than $\sqrt{\log (MK_0)/B}$ under alternative scenarios.

It is noteworthy that \cite{hu2021using} defined an alternative set $\{(K_0,g_0):K_0<K, \ell^{*}(K_0,g_0)/\sqrt{\log n}\to\infty\}$. By Condition (C3), we have that $\ell^{*}(K_0,g_0)/\sqrt{\log n}\to\infty$ implies $\sqrt{B}\max_vq_{v}/\sqrt{\log (MK_0)}\to \infty$. Therefore, in dense networks, $\{(K_0,g_0):K_0<K,\ell^{*}(K_0,g_0)/\sqrt{\log n}\to\infty\}$ is a subset of our alternative set $\mD(K,g,Q)$. Accordingly, our tests are not only robust against sparsity, but also powerful under \cite{hu2021using}'s alternative set.

To demonstrate the theoretical property of asymptotic power under  alternatives in the set $\mD(K,g,Q)$, we introduce the following condition on $K_0$ and $g_0$.
\begin{itemize}
\item [(C2*)] There exist finite constants $\xi_1^*>0$ and $\xi_2^*>0$ such that
$$\xi_1^* n/K_0\le\min_{1\le u\le K_0}|g_0^{-1}(u)|\le\max_{1\le u\le K_0}|g_0^{-1}(u)|\le\xi_2^* n^2/(K_0^2\log^2n).$$
\end{itemize}
This condition is mild and analogous to Condition (C2). Basically, it imposes bounds on community sizes for the hypothetical number of communities $K_0$ and the hypothetical membership vector $g_0$.
Consider the test statistic $\Theta_{n,2}=\Gamma_n^2(K_0,g_0)-2\log(MK_0)+\log\log(MK_0)$  defined in Subsection 2.3.
We next demonstrate its asymptotic property under  alternatives.

\bet\label{them-power}
Assume Conditions (C1), (C2*) and (C3) hold. We have that, for any alternative $(K_0, g_0)\in\mD(K, g, Q)$ and for any finite positive constant $c_1$, as $n\to\infty$,
\[P\Big\{\Theta_{n,2}\ge c_1\log (MK_0)\Big\}\to 1.\]
\eet
Theorem~\ref{them-power} indicates that $\Theta_{n,2}$ diverges at a rate of $\log (MK_0)$ under the alternative hypothesis in (II) that $(K_0,g_0)\in\mD(K,g,Q)$.
This result is analogous to Theorem 2 in \cite{hu2021using}. The main difference is
that our proposed $Q$ and $Q^{g_0}$ can decay to zero in sparse networks under the alternative set $\mD(K,g,Q)$. Accordingly, the resulting test statistic is applicable for both sparse and dense networks.
It is worth noting that Theorem 2 can also be used to calculate the power of $\Theta_{n,1}$ under the alternative hypothesis in (I).

Under the null hypothesis in (II), Theorem 1 shows $\Theta_{n,2}=O_p(1)$.
In contrast, under alternatives, Theorem 2 indicates that the divergence rate of $\Theta_{n,2}$ is $\log(MK_0)$.
Then, applying Condition (C3)-(iii), $MK=O(n^{\xi_4})$
with $0<\xi_4<1$, the rate of $\Theta_{n,2}$ is $\log(n)$, which is the same as that of \cite{hu2021using}.
From a theoretical perspective, the test statistic $\Theta_{n,2}$,
with  divergence  rate $\log(n)$, is sufficient to separate the null and alternatives asymptotically.
In addition, under Conditions (C3)-(ii),  $K=o\{(n/\log n)^{\xi_3}\}$ with $0<\xi_3<1/2$, and (C3)-(iii), $MK=O(n^{\xi_4})$
with $0<\xi_4<1$, we have that $M=O(n^{\xi_4})$.
As a result,  $\Theta_{n,2}$ is asymptotically powerful against the alternative set $\mD(K,g,Q)$, even for moderately large $M$.

It is noteworthy that we have provided three sufficient conditions to ensure that
the alternative $(K_0, g_0)$ is contained in $\mD(K,g,Q)$; see Proposition 1 in Section S.4 and
its proof in Sections S.5 of the supplementary material. In addition,
we have rigorously established the power function of $\Theta_{n,2}$ under some special alternatives in the set $D(K, g, Q)$, and found that the power of $\Theta_{n,2}$ is   expected to increase as the network becomes
denser (i.e., $Q$ increases), the network size $n$ grows, or the number of communities
$K$ decreases; see Proposition 2 and its related discussion in Section S.4, and
the proof in Section S.6 of the supplementary material.

\subsection{Bootstrap-corrected test statistic}

Note that the speed at which $\Theta_{n,1}$ or $\Theta_{n,2}$ converges to the extreme distribution may be slow. We next
consider a finite-sample bootstrap-corrected test statistic by generating bootstrap samples to improve the finite sample performance.
This bootstrap correction was proposed by \cite{bickel2016hypothesis} and also
considered by \cite{lei2016goodness} and \cite{hu2021using}.
In the remainder of this article, we use generic notation $\Theta_n$ to denote either $\Theta_{n,1}$ or $\Theta_{n,2}$ for testing  (I) or (II).
Given an adjacency matrix $A$, $K=K_0$ and $g=g_0$, the bootstrap-corrected test statistic $\Theta_{n,boot}$ can be obtained using Algorithm 2 below to test (II),
which consists of four steps. Note that $g_0$ can be replaced by its consistent
estimator $\hat g$ for testing (I).

\begin{algorithm}[htbp]
  \KwInput{Adjacency matrix $A$, the mapping vector $g_0$,
the number of communities $K_0$  produced by its corresponding $g_0$, the number of realizations $M$,
  the number of random samples $B$, the number of bootstraps $J$, and the significance level $\alpha$.}
  \KwOutput{Bootstrap-corrected test statistic $\Theta_{n,boot}$, and the decision in regard to the  null hypothesis being rejected or not. }
 Calculate $\hat Q$ and then obtain $\Theta_n$ from Algorithm 1 using $A$, $g_0$, $K_0$, $B$ and $M$.

  For $j=1,\cdots,J$, randomly generate a network $A^{(j)}$ based on the stochastic block model $(\hat Q, g_0)$, and obtain $\Theta^{(j)}_n$ from Algorithm 1 using $A^{(j)}$, $(\hat Q, g_0)$, $K_0$, $B$ and $M$.

  Estimate the location and scale parameters $\hat\mu$ and $\hat\beta$ of the Type-I extreme distribution using the maximum likelihood estimation method given $\Theta^{(j)}_n$s for $j=1,\cdots,J$.

  Calculate the bootstrap-corrected test statistic $\Theta_{n,boot} = \mu+\beta(\frac{\Theta_n-\hat\mu}{\hat\beta})$, where $\mu = -2\log(\sqrt{\pi})$ and $\beta=2$. Reject the null hypothesis if $\Theta_{n,boot}>q_{\alpha}$.
\caption{Bootstrap-Corrected Test Statistic}
\end{algorithm}

We can demonstrate theoretically that the bootstrap-corrected test statistic $\Theta_{n,boot}$ has the same asymptotic null distribution as
$\Theta_n$; see Corollary S.2 and its proof in Section S.7 of the supplementary material.
To examine the finite sample performance of these two tests, simulation studies  are reported in the next section. They indicate that
$\Theta_{n,boot}$ performs better than $\Theta_n$ in both dense and sparse networks.

\subsection{An augmented test}

As noted by \cite{hu2021using},
the power of their proposed test for hypothesis test (I) may be small when the networks are partitioned equally. To improve the power,
they proposed a novel technique of adding a new community to the observed network.
Consider a planted partition model from an adjacency matrix $A$ with $n$ nodes and $K$ equal-sized communities. Let $K^{+}_0 = K_0+1$.
Inspired by the approach of \cite{hu2021using},
we propose an augmented test statistic to test (I).
In  Algorithm 3 below, we first obtain the test statistic $\Theta^{+}_{n}$  to test (II), which consists of seven steps. Then we replace $g_0$ in $\Theta^{+}_{n}$ with its consistent
estimator $\hat g$ so that it can be used to test (I).

\begin{algorithm}[htbp]
  \KwInput{Adjacency matrix $A$, the mapping vector $g_0$, the number of communities $K_0$ produced by its corresponding $g_0$, the number of realizations $M$, the number of random samples $B$, and the significance level $\alpha$.}
  \KwOutput{Augmented test statistic $\Theta^{+}_{n}$ and the decision in regard to the  null hypothesis being rejected or not.}

Calculate the probability matrix $\hat Q$ using  (\ref{eq:hatB}).

  Add a $K^{+}_0$-th community of size $n_{K_0^{+}}=\min_{1\le u\le K_0}|g^{-1}_0(u)|/2$ to the observed network with $Q_{K_0^{+}K_0^{+}}= \max_{1\le u\le K_0} \hat Q^{g_0}_{uu}$ and $Q_{uK_0^{+}}=\min_{u\not=v} \hat Q^{g_0}_{uv}/2$ for $1\le u,v\le K_0$.

  Calculate the size $n^{+}=n+n_{K_0^{+}}$ and the adjacency matrix $A^{+}$ induced by the $n^+$ nodes from Step 2, and then calculate the probability matrix $\hat Q^{+}$ with the membership vector $g^{+}_0=(g_0, \underbrace{K^{+}_0,\cdots, K^{+}_0}_{n_{K^{+}_0}})$.

  Calculate $\hat\rho^{+}_{iv}$  using (\ref{eq:Zhang.hatrho}) with $A^{+}$, $g^{+}_0$ and $\hat Q^{+}$, for each $1\le i\le n^{+}$ and $1\le v\le K^{+}_0$.

At the  $m$-th realization for $m=1,\cdots M$, randomly draw $B$ samples from $\{\hat\rho^{+}_{1v},\cdots,\hat\rho^{+}_{n^{+}v}\}$ obtained in Step 4 for each $1\le v\le K^{+}_0$, and calculate $\hat\psi^{+}_{mv}$ using (\ref{eq:hat.gamma}).

Find the maximum value of $\hat\psi^{+}_{mv}$ for $1\le m\le M$ and $1\le v\le K^{+}_0$ and obtain $\Gamma_n(K^{+}_0,g_0^{+})$ using (\ref{eq:F0}).

 Calculate augmented test statistic $\Theta^{+}_{n}=\Gamma^2_n(K^{+}_0, g_0^{+})-2\log(MK^{+}_0)+2\log\log(MK^{+}_0)$ and reject the null hypothesis if $\Theta^{+}_{n}>q_{\alpha}$.
\caption{Augmented Test Statistic}
\end{algorithm}

It is worth noting that the above augmented procedure in Step 2 yields a larger disparity measurement, denoted by $q_{vu^*}$ in equation \eqref{eq:q(v)}.
As a result, $\sqrt{B} \max_{1 \le v \le K_0} q_{v}/\sqrt{\log(MK_0)}$ becomes larger, which
leads to a larger value of $\Theta^+_{n}$ so that its power is improved.
Additionally, we can demonstrate theoretically
that the asymptotic null distribution of $\Theta^{+}_{n}$ is identical to that of $\Theta_{n}$; see Corollary S.3 in Section S.7 of the supplementary material.
Moreover, to improve the power of the test statistic by controlling the type-I error close to the 5\% nominal level, we can integrate the bootstrap-corrected method introduced in Subsection 2.5 with this augmented test statistic and obtain the augmented bootstrap-corrected statistic $\Theta^{+}_{n,boot}$.
Analogously, one can verify that the asymptotic null distribution of $\Theta^{+}_{n,boot}$ is the same as that of $\Theta_n$.
According to the simulation results, $\Theta^{+}_{n,boot}$  can not only control the size well, but also improve the power.
Hence, we include $\Theta^{+}_{n,boot}$ in our simulation studies.


\section{Simulation studies}
\label{sec:verify}

To demonstrate the performance of the proposed test statistic, we conduct simulation studies testing the null hypotheses in (I) and (II) for both dense and sparse networks.
To test (I), we employ the simple spectral clustering algorithm  to estimate the community membership $\hat g$; see also \cite{lei2015consistency}. Specifically,
for the given adjacency matrix $A$ with $K_0$ hypothetical communities, one can recover the community membership $\hat g$ by applying the  $k$-means
clustering method to the rows of the first $K_0$ leading singular vectors of the adjacency matrix $A$.
In addition,  the network density in our simulation studies is $\tilde Q=\sum_{i,j}A_{i,j}/(n\times(n-1))$, where $n$ is the size of network.
Then, we set $B=\tilde Q^{-1/2}\{n/\log n\}^{1/3}$ and $M=100$. One can easily verify that these definitions of $B$ and $M$ satisfy Condition (C3).
The term $\tilde Q^{-1/2}$ in $B$ plays a role such that
the sparser the network, the more  samples are needed in (\ref{eq:hat.gamma}) to alleviate the sparsity.
Our simulation results below indicate that the selected $B$ and $M$ work satisfactorily for both dense and sparse networks.

In studying hypothesis test (I), we not only consider our  method, but also include
the four competing methods proposed by \cite{hu2021using}, \cite{han2023universal}, \cite{zhang2020adjusted}, and \cite{lei2016goodness}.
After including the augmented or the bootstrap-corrected
procedure applied to our method, \cite{hu2021using}'s method, and \cite{lei2016goodness}'s method, we include a total of eight tests  in our simulation studies for testing (I).
They are $\Theta_n$,  $\Theta^{+}_{n,boot}$, $\Theta_{Hu}$,
$\Theta^{+}_{Hu,boot}$, $\Theta_{Lei}$, $\Theta_{Lei,boot}$,
$\Theta_{Han}$, and $\Theta_{Zhang}$, where $\Theta_{Hu}$ and
$\Theta^{+}_{Hu,boot}$ are the test statistics proposed by
\cite{hu2021using}. It is worth noting that the augmented procedure is proposed to test (I),
and the methods of \cite{lei2016goodness},
\cite{han2023universal}, and \cite{zhang2020adjusted} are not applicable to test (II). Accordingly, we only consider the test statistics $\Theta_n$ and $\Theta_{Hu}$ in testing  (II).
Due to space constraint, all the simulation results for hypothesis test (II) are given in the supplementary material;
see Tables S.1-S.2 of Section S.8. In simulation studies, all simulation results are evaluated using 500 realizations.

\subsection{Simulations under dense networks}
In this subsection,
we modify the setting specified by Hu et al. (2021) to construct two dense networks to test hypothesis (I), namely Setting 1.

$\bm{Setting}$ $\bm{1:}$
We set the edge probabilities between any two communities $u$ and $v$ to be $0.1(1 + 4\times I(u = v))$.
As a result, $Q_{g(i)g(j)}=0.1$ if $g(i)\not=g(j)$ and $Q_{g(i)g(j)}=0.5$ otherwise.
Thus, $\min_{i,j} Q_{g(i)g(j)}=0.1$, the edge probabilities
do not decay with $n$, and the resulting network is a typical dense network.
Then, we consider the following two cases: (i) the size of each community $n_k=300$ with $K$ communities for $k=1,\cdots, K$,
the total number of nodes  in this
network is $n=Kn_k$, and $K_0$= 2, 4, 6, 8, and 10;
(ii) the size of the network is $n=3,000$ with $K$ blocks, the size of each block is $3000/K$, and
$K_0$= 3, 5, 10, 15, and 20.

Tables 1-2 present the sizes and powers of all eight tests
for cases (i) and (ii), respectively.
They show the following findings.
First, our proposed tests $\Theta_n$ and  $\Theta^{+}_{n,boot}$  generally perform better than $\Theta_{Hu}$ and $\Theta^{+}_{Hu,boot}$ in controlling the size, and the sizes of $\Theta_{Hu}$ and $\Theta^{+}_{Hu,boot}$ are significantly and slightly distorted, respectively, when $K > 10$.
 This finding is reasonable, because the sampling process can mitigate
 the instability
of the spectral clustering estimate $\hat g$. Recall that our proposed statistic is constructed by sampling
the $B$ entry-wise deviations obtained from \cite{hu2021using}. Hence, as $B\rightarrow \infty$, the negative impact due to
a small number of poorly estimated $\hat g$ can be ignored.
Thus, the performance of the proposed
statistic can be dominated by a large number of consistently estimated membership vectors $\hat g$.
In addition,  both $\Theta_n$ and  $\Theta^{+}_{n,boot}$ outperform $\Theta_{Lei}$, $\Theta_{Lei,boot}$, $\Theta_{Han}$ and $\Theta_{Zhang}$ in controlling the size.  Specifically, the latter four methods exhibit significant size distortions when $K\ge 8$, since their empirical sizes are significantly larger than the nominal level of 5\%. Consequently, $\Theta^{+}_{n,boot}$ and $\Theta_n$ perform the best overall for controlling the size  in dense networks.

Second, both $\Theta_n$ and $\Theta_{Hu}$ exhibit weaker power against alternatives than those of $\Theta_{Lei}$, $\Theta_{Lei,boot}$, $\Theta_{Han}$, and $\Theta_{Zhang}$ in dense networks, and $\Theta_n$ is also less powerful than $\Theta_{Hu}$.
As expected, the sizes and powers of $\Theta_n$ and $\Theta_{Hu}$ can be enhanced by $\Theta^{+}_{n,boot}$ and $\Theta^{+}_{Hu,boot}$, respectively, since they
integrate the bootstrap-corrected method with the augmentation procedure. Tables 1-2 demonstrate that the empirical powers of $\Theta^{+}_{n,boot}$ and $\Theta^{+}_{Hu,boot}$ are comparable to those
of $\Theta_{Lei}$, $\Theta_{Lei,boot}$, and $\Theta_{Zhang}$, and stronger than that of $\Theta_{Han}$.
In addition,
$\Theta^{+}_{n,boot}$ is less powerful than $\Theta^{+}_{Hu,boot}$, in particular when $K$ is large, as seen in Table 2.
This finding is not surprising since the size of $\Theta^{+}_{Hu,boot}$ is distorted for large $K$. Overall, $\Theta^{+}_{n,boot}$ is a powerful test.

\makeatletter\def\@captype{table}\makeatother
\begin{spacing}{1}
\setlength{\abovecaptionskip}{0.5cm}
\setlength{\belowcaptionskip}{0.2cm}
\scriptsize
\centering
\caption{ Setting 1 (i): Proportion of rejections at nominal level $\alpha = 0.05$ for hypothesis test $H_0 : K = K_0$ v.s.
$H_1 : K > K_0$. Each community has 300 nodes and $Q_{uv} = 0.1(1 + 4 \times 1(u = v))$.}
\label{Tab:01}
\renewcommand\arraystretch{1.5}{
\begin{tabular}{c|ccccc|ccccc}
\hline
& \multicolumn{5}{c|}{$\Theta_{n}$} &\multicolumn{5}{c}{$\Theta^{+}_{n,boot}$} \\
 \cline{1-11}
 K &2 & 4& 6& 8 & 10 &2 & 4& 6& 8 & 10 \\
\hline
$K_0=2$ &  0.035& 0.105& 0.225& 0.425& 0.570&0.045& 1& 1& 1& 1\\
$K_0=4$ &*&0.040& 0.095& 0.140& 0.335&*&0.045&  1& 1&  1 \\
$K_0=6$ &* &* &0.065& 0.090& 0.170& * &* &0.065&  0.975& 1\\
$K_0=8$ &* &* &*& 0.035&  0.065&* &* &* &0.055&   0.840\\
$K_0=10$ &* &* &* &* &0.050 &*	&* &* &* &0.050\\
\hline
&\multicolumn{5}{c|}{$\Theta_{Hu}$}& \multicolumn{5}{c}{$\Theta^{+}_{Hu,boot}$}\\
 \cline{1-11}
 K &2 & 4& 6& 8 & 10 &2 & 4& 6& 8 & 10\\
\hline
$K_0=2$ & 0.025& 0.100& 0.385& 0.640& 0.840&  0.055& 0.995& 1& 1& 1\\
$K_0=4$ &* &0.040& 0.195& 0.410& 0.675 &*&0.045&  1& 1&  1 \\
$K_0=6$ &* &* &0.055& 0.185& 0.385&* &* &0.075&  1& 1\\
$K_0=8$ &* &* &* &0.065&   0.160&* &* &* &0.055&  0.995\\
$K_0=10$ &*	&* &* &* &0.070 &* &* &* &*	&0.065 \\
\hline
&\multicolumn{5}{c|}{$\Theta_{Lei}$}& \multicolumn{5}{c}{$\Theta_{Lei,boot}$}\\
 \cline{1-11}
 K &2 & 4& 6& 8 & 10 &2 & 4& 6& 8 & 10\\
\hline
$K_0=2$ & 0.050&  1&   1&   1& 1&  0.045& 1&   1&  1&1 \\
$K_0=4$ & *& 0.080&   1&   1& 1 & *& 0.065& 1&  1&1\\
$K_0=6$ & *& *& 0.115&  1& 1&  *& *& 0.080&  1& 1\\
$K_0=8$ & *& *& *& 0.170& 1&  *& *& *& 0.135&1 \\
$K_0=10$ & *&  *& *& *&0.305 &  *& *& *& *& 0.290\\
\hline
&\multicolumn{5}{c|}{$\Theta_{Zhang}$} &\multicolumn{5}{c}{$\Theta_{Han}$}\\
 \cline{1-11}
 K &2 & 4& 6& 8 & 10 &2 & 4& 6& 8 & 10\\
\hline
$K_0=2$ & 0.075&   1&  1&  1&  1& 0.040& 0.715& 0.820&  0.790&0.815 \\
$K_0=4$ & *& 0.040&  1&  1& 1& *& 0.065& 1& 1&0.995\\
$K_0=6$ & *& *& 0.090& 0.995& 0.995& *& *& 0.065&  1&  1\\
$K_0=8$ & *& *&  *&  0.100& 0.670&  *& *& *& 0.095&  1\\
$K_0=10$ & *& *& *& *& 0.180& *& *&  *& *& 0.250\\
\bottomrule
\end{tabular}}
\end{spacing}

\newpage

\makeatletter\def\@captype{table}\makeatother

\begin{spacing}{1}
\setlength{\abovecaptionskip}{0.5cm}
\setlength{\belowcaptionskip}{0.2cm}
\scriptsize
\centering

\caption{ Setting 1 (ii): Proportion of rejections at nominal level $\alpha = 0.05$ for hypothesis test $H_0 : K = K_0$ v.s.
$H_1 : K > K_0$. The network size is $n = 3,000$ with equal sized communities, and $Q_{uv} = 0.1(1+4\times1(u = v))$.}

\label{Tab:02}
\renewcommand\arraystretch{1.5}{
\begin{tabular}{c|ccccc|ccccc}
\hline
& \multicolumn{5}{c|}{$\Theta_{n}$} &\multicolumn{5}{c}{$\Theta^{+}_{n,boot}$}\\
 \cline{1-11}
 K &3 & 5& 10& 15 & 20 &3 & 5& 10& 15 & 20\\
\hline
$K_0=3$ &   0.035& 0.090&  0.470& 0.730& 0.780&  0.030&  1&  1&  1& 1\\
$K_0=5$ & * &0.035&  0.145&  0.490&  0.660&* &0.055&  1&  1& 1\\
$K_0=10$ &* &* &0.050& 0.120&  0.235&* &* &0.045&  0.645&  0.715\\
$K_0=15$ &*	&* &* &0.035& 0.125&*	&* &* &0.055&  0.315\\
$K_0=20$ &* &* &* &* &0.040	&* &* &* &*	&0.040\\
\hline
&\multicolumn{5}{c|}{$\Theta_{Hu}$}& \multicolumn{5}{c}{$\Theta^{+}_{Hu,boot}$}\\
 \cline{1-11}
 K &3 & 5& 10& 15 & 20 &3 & 5& 10& 15 & 20\\
\hline
$K_0=3$ &0.020& 0.230& 0.865&  1&   1&   0.030&  1& 1&   1&  1\\
$K_0=5$ & * &0.025&  0.500& 0.995&  1 &* & 0.040&  1&  1& 1\\
$K_0=10$ &*&* &0.070&  0.400& 0.980&* &* &0.050& 0.990&  1 \\
$K_0=15$ &*	&* &* &0.140&  0.360&*	&* &* &0.090&   0.895\\
$K_0=20$ &* &* &* &* &0.235 &* &* &* &* &0.080 \\
\hline
 &\multicolumn{5}{c|}{$\Theta_{Lei}$}& \multicolumn{5}{c}{$\Theta_{Lei,boot}$}\\
 \cline{1-11}
 K  &3 & 5& 10& 15 & 20 &3 & 5& 10& 15 & 20\\
\hline
$K_0=3$ &0.065&1 &1 &1 &1 &  0.070&1 & 1&1 &1 \\
$K_0=5$ & *& 0.060& 1 &1 &1 & *& 0.065& 1& 1&1 \\
$K_0=10$ &* &*&0.245&1&1 &* &*&0.235&1& 1\\
$K_0=15$  &* &* &*&0.630& 1&* &* &* &0.615& 1\\
$K_0=20$  &*&*&*&*&0.725&*&*&*&*&0.925\\
\hline
 &\multicolumn{5}{c|}{$\Theta_{Zhang}$}& \multicolumn{5}{c}{$\Theta_{Han}$}\\
 \cline{1-11}
 K  &3 & 5& 10& 15 & 20 &3 & 5& 10& 15 & 20 \\
\hline
$K_0=3$ &0.105& 1& 1& 1& 1 & 0.060&1&0.955&0.875&0.750\\
$K_0=5$ & *& 0.070& 1& 1&  1 & *&0.070&1&0.975&0.955\\
$K_0=10$ &*&*& 0.180& 0.810& 0.915&* &* &0.220& 1&1\\
$K_0=15$  &*&*&*&0.150& 0.830&* &* &*&0.625&1\\
$K_0=20$  &*&*&*&*&0.240&*&*&*&*&0.690\\
\bottomrule
\end{tabular}}
\end{spacing}

~\\

In sum, Tables 1-2 demonstrate that
$\Theta^{+}_{n,boot}$ performs the best overall as compared with  other
tests in testing (I), and $\Theta_n$ controls size well.

\subsection{Simulations under sparse networks}

In this subsection, we consider the sparse network setting to test (I), namely Setting 2.

$\bm{Setting}$ $\bm{2:}$ We construct two types of sparse networks as follows: (i)
the edge probabilities between any two communities $u$ and $v$ are $2\log\log n(1 + 4\times I(u = v))/n$.
The size of each block is $n_k=300$, for $k=1,\cdots, K$, the total number of nodes in the network is $n=Kn_k$, and $K_0$= 2, 4, 6, 8, and 10; (ii)
the edge probabilities between any two communities $u$ and $v$ are $3\log\log n(1 + 4\times I(u = v))/n$, and the size of the
network is $n=3,000$ with $K$ blocks. The network size of each block is $3000/K$, and $K_0$= 2, 3, 4, 5, 6, 8, and 10.
The above two settings lead to
 $\max_{i,j} Q_{g(i)g(j)}=10 \log\log n/n$ and $\max_{i,j} Q_{g(i)g(j)}=15 \log\log n/n$
for networks (i) and (ii), respectively.
Accordingly, $\max_{i,j} Q_{g(i)g(j)}=O(n^{-1}\log\log n)$, and it typically yields
sparse networks.

Based on the network density conditions imposed by \cite{lei2016goodness}, \cite{hu2021using}, \cite{zhang2020adjusted}, and \cite{han2023universal}, their resulting test statistics, $\Theta_{Lei}$, $\Theta_{Lei,boot}$, $\Theta_{Hu}$, $\Theta^{+}_{Hu,boot}$, $\Theta_{Zhang}$, and $\Theta_{Han}$, are not
suitable for networks with sparse density on the order of $O(n^{-1}\log\log n)$.
In addition, the methods proposed by \cite{zhang2020adjusted} and \cite{han2023universal} required a fixed value of $K$.  Hence, their test statistics may not be
applicable to data with a large $K$. Our simulation
results in Tables 3--4 corroborate these assertions. Specifically, the empirical sizes
of $\Theta_{Lei}$, $\Theta_{Hu}$, and $\Theta^{+}_{Hu,boot}$ are all significantly greater than 5\% and suffer from size distortions.
Additionally, the empirical sizes of $\Theta_{Zhang}$ are somewhat erratic and none of them approximate to the 5\% significance level. As for the tests $\Theta_{Han}$
and $\Theta_{Lei,boot}$, they perform poorly for $K\ge 3$, exhibiting  size distortions. On the other hand,
the sizes of $\Theta_n$ and $\Theta^{+}_{n,boot}$ are close to the nominal level of 5\%.
Since the sizes of the six competing tests are distorted, it is not sensible to make power comparisons.
In sum, under sparse networks, Tables 3-4 indicate that our proposed tests, $\Theta_n$ and $\Theta^{+}_{n,boot}$, perform  well in comparison with the other six tests, and $\Theta^{+}_{n,boot}$ is more powerful than $\Theta_n$, which supports our theoretical results.

Based on Tables 1-4, we conclude that our proposed test $\Theta^{+}_{n,boot}$ performs the best overall in testing (I) under both dense and sparse networks, and $\Theta_n$ can control empirical sizes well.
In addition, \cite{hu2021using}'s test, $\Theta^{+}_{Hu,boot}$, can be used for dense networks when $K$ is not large.

\begin{spacing}{1}
\setlength{\abovecaptionskip}{0.5cm}
\setlength{\belowcaptionskip}{0.2cm}
\scriptsize
\centering
\caption{ Setting 2 (i): Proportion of rejections at nominal level $\alpha = 0.05$ for hypothesis test $H_0 : K = K_0$ v.s.
$H_1 : K > K_0$. Each community has 300 nodes and $Q_{uv} = 2\log\log n(1 + 4 \times 1(u = v))/n$.}
\label{Tab:05}
\renewcommand\arraystretch{1.5}{
\begin{tabular}{c|ccccc|ccccc}
\hline
& \multicolumn{5}{c|}{$\Theta_{n}$} &\multicolumn{5}{c}{$\Theta^{+}_{n,boot}$}\\
 \cline{1-11}
 K &2 & 4& 6& 8 & 10    &2 & 4& 6& 8 & 10\\
\hline
$K_0=2$ &0.033  &0.075&0.140 &0.195  &0.245 &0.060 &0.320&0.515&0.580 &0.620\\
$K_0=4$ &* &0.036 &0.085 &0.155 &0.195 &* & 0.043&0.380&0.430&0.500\\
$K_0=6$ &* &*&0.045 &0.090 &0.180 &*  &*  &0.055 &0.235 &0.392 \\
$K_0=8$ &* &* &* &0.045 &0.140 &* &* &* &0.060 &0.220\\
$K_0=10$  &* &* &*	&* &0.065 &* &* &*	&* &0.060\\
\hline
&\multicolumn{5}{c|}{$\Theta_{Hu}$}& \multicolumn{5}{c}{$\Theta^{+}_{Hu,boot}$}\\
 \cline{1-11}
 K &2 & 4& 6& 8 & 10 &2 & 4& 6& 8 & 10\\
\hline
$K_0=2$ &0.610  &0.910 &0.990 &0.990&1 &0.970 &0.900&0.975&0.985&0.980\\
$K_0=4$ &* &1 & 1&1 &1  &* & 0.900&0.990&0.995&0.995\\
$K_0=6$ &* &* &1 &1  &1  &* &* &0.990 &1  &1 \\
$K_0=8$ &* &* &* &1 &1  &* &* &* &0.996 &1\\
$K_0=10$ &* &* &* &* & 1  &* &* &* &* &1 \\
\hline
&\multicolumn{5}{c|}{$\Theta_{Lei}$}& \multicolumn{5}{c}{$\Theta_{Lei,boot}$}\\
 \cline{1-11}
 K  &2 & 4& 6& 8 & 10   &2 & 4& 6& 8 & 10\\
\hline
$K_0=2$ &1 &1 &1  &1  &1  &0.050  &1  &0.265 & 0.487 &1 \\
$K_0=4$  & *&1 &1  &1  &1    &*&0.025 &1&0.337  &0.644  \\
$K_0=6$ & *& *&1  &1  &1 & *& *&0.226&0.500&0.556   \\
$K_0=8$ & *&*&*&1 &1 & *&*&*&0.274 &0.543 \\
$K_0=10$ & *& *& *& *&1  & *& *& *& *&0.221\\
\hline
&\multicolumn{5}{c|}{$\Theta_{Zhang}$}& \multicolumn{5}{c}{$\Theta_{Han}$}\\
 \cline{1-11}
 K  &2 & 4& 6& 8 & 10   &2 & 4& 6& 8 & 10 \\
\hline
$K_0=2$ &0.080 & 0.980 &0.210 &0.165  &0.125     &0.030 &0.745 &0.910 &0.950 &0.315   \\
$K_0=4$  &*&0.125 &0.115 &  0.105&0.160    &* &0.635&0.830&0.980&0.955 \\
$K_0=6$  & *& *&0.020 & 0.065 &0.080     &*& *&0.960 &0.930&0.935    \\
$K_0=8$ & *& *& *&0.008&0.070     & *& *& * & 0.930&0.875\\
$K_0=10$ & *& *& *& *&0.003   & *& *& *& *&0.590 \\
\bottomrule
\end{tabular}}
\end{spacing}

\newpage

\makeatletter\def\@captype{table}\makeatother
\begin{spacing}{1}
\setlength{\abovecaptionskip}{0.5cm}
\setlength{\belowcaptionskip}{0.2cm}
\scriptsize
\centering
\caption{ Setting 2 (ii): Proportion of rejections at nominal level $\alpha = 0.05$ for hypothesis test $H_0 : K = K_0$ v.s.
$H_1 : K > K_0$. The network size is $n = 3,000$ with equal sized communities, and $Q_{uv} = 3\log\log n(1+4\times1(u = v))/n$.}
\label{Tab:06}
\renewcommand\arraystretch{1.45}{
\begin{tabular}{c|ccccccc|ccccccc}
\hline
& \multicolumn{7}{c|}{$\Theta_{n}$} &\multicolumn{7}{c}{$\Theta^{+}_{n,boot}$}\\
 \cline{1-15}
 K &2 & 3& 4& 5 & 6 &8 &10   &2 & 3& 4& 5 & 6 &8 &10\\
\hline
$K_0=2$ &0.043&0.060 &0.110 &0.170 &0.265 &0.325&0.360     &0.060&0.650 &0.685 &0.750&0.780 &0.800&0.825\\
$K_0=3$ &* & 0.055&0.095 &0.115 &0.135 &0.205 &0.320&* & 0.050&0.560&0.685&0.700&0.765&0.800\\
$K_0=4$ &* &* & 0.055 &0.065&0.130&0.185 &0.210 & * &* & 0.060&0.420&0.485&0.565&0.620\\
$K_0=5$ &* &* &* &0.065 &0.110 &0.160 &0.195 & * &* &* &0.040 &0.365&0.455&0.550\\
$K_0=6$ &* &* &* &*  &0.070 &0.105 &0.160 &* &* &* &*  &0.035 &0.305 &0.355 \\
$K_0=8$ &* &* &* &*  &* &0.065 &0.110 &* &* &* &*  &* &0.070 &0.310 \\
$K_0=10$ &* &* &*	&* &* &* & 0.070 &* &* &*  &* &* &* & 0.060\\
\hline
& \multicolumn{7}{c|}{$\Theta_{Hu}$}& \multicolumn{7}{c}{$\Theta^{+}_{Hu,boot}$} \\
 \cline{1-15}
 K &2 & 3& 4& 5 & 6 &8 &10   &2 & 3& 4& 5 & 6 &8 &10\\
\hline
$K_0=2$ &0.537&0.960&0.970&0.985&0.995&1 &1 &0.993&0.930&0.950&0.970&0.995&1&1 \\
$K_0=3$ &* & 0.975 &1&1 &1  &1 &1  &* & 0.870 &0.980&0.980&0.995&1&1\\
$K_0=4$ &* &* & 1 & 1&1 & 1&1  & * &* & 0.815 &0.975&1&1&1\\
$K_0=5$ &* &* &* &1  &1 &1 & 1 & * &* &* &0.955 &1&1&1\\
$K_0=6$ &* &* &* &*  &1  &1  &1  &* &* &* &*  &1  &1 &1\\
$K_0=8$ &* &* &* &*  &* & 1  & 1 &* &* &* &*  &* & 1 &1\\
$K_0=10$ &*  &* &* &* &*  &* & 1 &*  &* &* &* &*  &* & 1  \\
\hline
& \multicolumn{7}{c|}{$\Theta_{Lei}$}& \multicolumn{7}{c}{$\Theta_{Lei,boot}$}\\
 \cline{1-15}
 K &2 & 3& 4& 5 & 6 &8 &10   &2 & 3& 4& 5 & 6 &8 &10\\
\hline
$K_0=2$ &1&1 &1  &1  &1  &1  & 1 & 0.050&1 &1  &1  &0.495  &0.245  &0.517  \\
$K_0=3$ &*&1 &1  &1  &1  & 1 &1  &*&0.100 &1 &1  &0.140  &0.368 &0.614 \\
$K_0=4$ &*&*&1&1  &1  &1  &1   &*&*&0.080 &1 &0.250  &0.166  &0.399  \\
$K_0=5$  &*&*& *&1 &1  &1 &1 &*&*& *&0.110 &1&0.550&0.156 \\
$K_0=6$ &* &* &*&* &1 &1 &1  &* &* &*&* &0.165 &0.518&0.477 \\
$K_0=8$ &* &* &* &* &* &1 &1 &* &* &* &* &* &0.388 &0.473 \\
$K_0=10$ &*  &* &* &* &*  &* &1 &*  &* &* &* &*  &* &0.343 \\
\hline
&\multicolumn{7}{c|}{$\Theta_{Zhang}$}& \multicolumn{7}{c}{$\Theta_{Han}$}\\
 \cline{1-15}
 K &2 & 3& 4& 5 & 6 &8 &10   &2 & 3& 4& 5 & 6 &8 &10\\
\hline
$K_0=2$ &0.130 &1&1&1&0.995&0.355&0.145    &0.025&0.795&0.760&0.620 &0.620 &0.525 &0.632\\
$K_0=3$ &*&0.075&1&1&0.995&0.267&0.313   &*&0.085&0.605&0.805 &0.770 &0.848 &0.900\\
$K_0=4$ &*&*&0.025&1&0.975&0.133&0.153    &*&*&0.150&0.745&0.817 &0.945 &0.985\\
$K_0=5$ &*&*&*&0.010&0.420&0.193&0.113    &*&*&*&0.360&0.910&0.940&0.960\\
$K_0=6$ &*&*&*&*&0.010&0.080&0.080 &*&*&*&*&0.865 &0.965 &0.965 \\
$K_0=8$ &*&*&*&*&*&0.030&0.070 &*&*&*&*&*&0.915 &0.930\\
$K_0=10$ &*&*&*&*&*&*&0.020 &*&*&*&*&*&*&0.890\\
\bottomrule
\end{tabular}}
\end{spacing}

~\\

\begin{remark}
	We also construct simulation studies under Settings 1 and 2 to test hypothesis (II).
	Due to space constraint, all of the simulation results for testing hypothesis (II) are given in Tables S.1-S.2 of Section S.8 in the supplementary material.
	The simulation results in Tables S.1-S.2 are consistent with those findings in Tables 1-4. Under Setting 1, Table S.1 indicates that
	our proposed statistic $\Theta_n$ is superior to $\Theta_{Hu}$ in the size control, and both $\Theta_n$ and $\Theta_{Hu}$ are powerful against the alternatives with the empirical powers all equal to 1.
	Under Setting 2, Table S.2 shows that the empirical sizes of $\Theta_n$ are close to the nominal level of 5\% and the empirical powers of $\Theta_n$ are all 1, while $\Theta_{Hu}$ fails to control size well.  In sum, when testing hypothesis (II), $\Theta_n$ performs well and is superior to $\Theta_{Hu}$ in the size control under both dense and sparse networks.
\end{remark}

\begin{remark}
	It is worth noting that the condition $D_n=O(n^{-1}\log n)$ has been considered in obtaining theoretical properties of various
	tests; see, e.g., the $\Theta_{Zhang}$ test. According to this condition, we construct additional simulation studies under Setting 2 to test hypotheses (I) and (II), except that the edge probabilities between any two communities $u$ and $v$ are adjusted to $2\log n(1 + 4\times I(u = v))/n$ and $3\log n(1 + 4\times I(u = v))/n$, respectively, for cases (i) and (ii).
	Additional simulation results under this setting are presented in Tables S.3-S.5 of Section S.9 in the supplementary material.
	The simulation results in Tables S.3-S.5 are consistent with those findings in Tables 3-4.
	In sum, when testing hypothesis (I), our proposed test statistics $\Theta_n$ and $\Theta^{+}_{n,boot}$ outperform all competing methods, and $\Theta^{+}_{n,boot}$ is more powerful than $\Theta_n$ and performs the best overall. When testing hypothesis (II), $\Theta_n$ is
	superior to $\Theta_{Hu}$.
	
\end{remark}

\section{Real data analyses}
\label{sec:conc}

To demonstrate the usefulness of our proposed method for both dense and sparse networks,
we analyze two real datasets. Based on simulation studies, we find that our proposed method and \cite{hu2021using}'s method
generally outperform those methods proposed by \cite{lei2016goodness}, \cite{han2023universal}, and \cite{zhang2020adjusted}. Thus, in our real data analyses, we only consider our proposed method and \cite{hu2021using}'s method.
The first dataset is the international trade data, which has been studied by \cite{westveld2011mixed}, \cite{lei2016goodness}, and \cite{hu2021using}.
This data set contains yearly trade information of 58 countries.
\cite{hu2021using} constructed a network
with density of 48.1\% (=$1590/(58\times 57)\times$ 100\%), and it is
a typical dense network.
As noticed by \cite{hu2021using}, the augmentation procedure may not work well for this  data set because the size of the network is small and the added community may have
less than or equal to two nodes.
Therefore, we apply the test statistics, $\Theta_{n,boot}$ and $\Theta_{Hu,boot}$, to analyze the international trade network.
The second data set is the co-authorship network dataset collected by \cite{ji2016coauthorship}.
This network has 3,607 nodes and the density of this network is approximately equal to 0.012\% (=$1561/(3607\times 3606)\times $100\%), which makes it a sparse network.
The simulation results indicate that $\Theta^{+}_{n,boot}$ and $\Theta^+_{Hu,boot}$ perform
 better
than $\Theta_n$ and $\Theta_{Hu}$, respectively.
Hence, we employ them to study the co-authorship network.
However, the augmented tests are not applicable for testing
$H_0: K_0=1$ since it is infeasible to specify the connection probability between communities.
Thus, $\Theta_{n,boot}$ and $\Theta_{Hu,boot}$ are considered in this case.
Finally, the number of bootstraps used in both examples is $M=100$, and the number of samples drawn from $n$ nodes
is $B=\tilde Q^{-1/2}\{n/\log n\}^{1/3}$ as given in simulation studies.

Based on the simulation results presented in Tables 1-4,
our proposed method not only is superior to or comparable with the methods proposed by Lei (2016), Hu et al. (2021),
Han et al. (2023) and Zhang and Amini (2023) in dense networks, but also outperforms them in sparse networks.
Hence, our method is robust against network density (or sparsity).
In empirical analyses, we consider two examples corresponding to dense and sparse networks. Based on theoretical and simulation results, our proposed methods
are applicable for the two data sets.

\subsection{Dense network case: International trade data}

This dataset contains yearly trade information for $n=58$ countries
from 1981 to 2000. For any given year, $\text{Trade}_{ij}$ represents the value of exports from country $i$ to country $j$;
further descriptions can be found in \cite{westveld2011mixed}.
Following  \cite{saldana2017many} and \cite{hu2021using}, we consider
the international trade data in 1995, and construct the network from the 58 countries based on $\text{Trade}_{ij}$ as follows.
Each node represents a country,
and define $W_{ij}=\text{Trade}_{ij}+\text{Trade}_{ji}$. Then, the adjacency matrix $A=(a_{ij})$
is defined such that $a_{ij}=1$ if $W_{ij}>W_{0.5}$ and $a_{ij}=0$ otherwise, where $W_{0.5}$ is the 50\% quantile of the $W_{ij}$ values.
As discussed in \cite{saldana2017many}, it is possible to decompose the network into 3, 7, and 10 communities by three different methods.
However, for different choice of $K$, the resulting communities from these 58 countries vary. Therefore, determining the
appropriate number of communities is critical to label communities and make meaningful interpretations for 58 countries.
To this end, we employ the statistics $\Theta_{n,boot}$ and $\Theta_{Hu,boot}$ to test the three hypotheses $H_0: K=3$,  $H_0: K=7$, and $H_0: K=10$, respectively.

Given the  5\% significance level,
we apply our Theorem 1 and \cite{hu2021using}'s Theorem 1
to obtain the critical values of  $\Theta_{n,boot}$ and $\Theta_{Hu,boot}$, which are
$t_{n,0.95}=4.792$ and $t_{Hu,0.95}=3.413$, respectively.
For $K_0=3, 7,$ and 10, \cite{hu2021using} calculated their corresponding test statistics, $\Theta_{Hu,boot}=44.28, 3.33$ and  13.44. Comparing them with the critical value, 3.413, they concluded that $K=7$ is a reasonable choice. We next compute our proposed test statistics, $\Theta_{n,boot}=15.02, 3.07,$ and 16.76, for $K_0=3, 7,$ and 10, respectively.
Comparing them with the critical value, 4.792, we find that $\Theta_{n,boot}$ does not reject the hypothesis $H_0:K=7$, whereas it  rejects the hypotheses $H_0:K=3$ and
$H_0:K=10$.
Hence, $\Theta_{n,boot}$
yields the same finding as that of \cite{hu2021using}.

Based on our test results, the 58 countries can be decomposed into seven blocks.
Accordingly, we apply the spectral clustering algorithm of Lei and Rinaldo (2015) and classify 58 countries into seven communities.
The results are depicted in Figure S.1 of the supplementary material,
which allows us to make the following interpretations regarding the seven communities.
The first community marked in purple
comprises 4 countries including Cyprus, Egypt and Greece, located in the Mediterranean.
The second community marked in blue-purple
comprises 6 countries including Argentina, Bolivia and Chile, located in South America.
The third community marked in green includes 3 countries, Algeria, Morocco and Tunisia, all located in North Africa.
The fourth community marked in blue-green consists of 12 countries including Italy, Spain and Portugal, and most of these countries are located in Europe.
It seems that these four communities are identified by regional location, as these countries may be focusing in part on a local trading bloc that provides preferential access.
The fifth community marked in blue comprises 9 countries including Norway, Singapore and Denmark,  most of which have a high average GDP.
The sixth community marked in yellow consists of 13 countries, including USA, Germany, and Japan, most of which are highly influential over global trading volume.
The seventh community marked in orange comprises 11 countries including Australia, Canada and the United Kingdom,
many of which serve as important hubs of global trade and have GDPs
exceeding the global average. The last three communities seem to be identified by these countries' GDP or trading volume; hence, these countries play a pivotal role in global commerce.
This decomposition can provide a better understanding of
supply and demand via international trade.

\subsection{Sparse network case: Co-authorship data}

We next analyze the co-authorship dataset collected by \cite{ji2016coauthorship}.
This dataset contains all research papers published from 2003 to the first half of 2012 in four of the top statistical journals: {\it Annals of Statistics} (AoS),
{\it Biometrika,} {\it the Journal of American Statistical Association} (JASA), and {\it the Journal of Royal Statistical Society (Series B)} (JRSS-B). The dataset contains $n=3,607$ authors. According to \cite{ji2016coauthorship}'s Coauthorship Network (A), we construct the network as follows. Each node represents a author.
For any author $i$ and author $j$ ($i\not=j$), we define $a_{ij}=1$ for $1\le i,j\le n$
if and only if they have co-authored at least two papers during the study period.

The density of this network is approximately 0.012\%, which is smaller than 0.23\% (i.e., $\log(n)/n$ with $n=3,607$). As mentioned at the beginning of Section 4,  this network is considered sparse.
Utilizing this network data, \cite{ji2016coauthorship} employed various community detection methods to identify meaningful communities or
research groups. Specifically, based on their result in Table 1, the dataset is comprised of eight
communities.
In their studies, the number of communities corresponds to the number of major research topics or research institutions.
Therefore, identifying the number of communities is crucial for their analyses.

Using this sparse network data, we  test the
 hypothetical number of communities $K_0\in[1,8]$.
Given the  5\% significance level,
we have
$t_{n,0.95}=4.792$ and $t_{Hu,0.95}=3.413$ as given in the
above example.
In testing $H_0: K=1$, both $\Theta_{n,boot}$ and $\Theta_{Hu,boot}$ are
larger than 400, which results in rejecting the null hypothesis.
For $K_0\in[2,8]$, we have $\Theta^+_{Hu,boot}>7$, which results in rejecting all of their corresponding  hypotheses.
This finding is not surprising since
$\Theta^+_{Hu,boot}$ fails to converge to an extreme-value distribution for sparse networks.
In testing $K_0\in[2,8]$, we calculate
$\Theta^{+}_{n,boot}$=14.81, 15.80, 4.23, -1.53, 3.75, -2.52, and 3.08.
As a result, $\Theta^{+}_{n,boot}$ does not reject each of those hypotheses from
$H_0:K=4$ to $H_0:K=8$ at the 5\% significance
level.
Hence, $\Theta^{+}_{n,boot}$ can identify  $K=4$ to $K=8$ communities for all authors.

Based on \cite{ji2016coauthorship}'s finding in their Table 1, $K=8$ is sensible.
As for $K=4$, we can identify  the following four communities in the Coauthorship Network (A): ``High-Dimensional Data'', ``Machine Learning'',
``Experimental Design'' and ``Prestige and Non-Ivy League Universities''.
Please note that the community of ``High-Dimensional Data''  and ``Machine Learning'' are merged from  ``High-Dimensional Data Analysis and Dimension Reduction'' and ``Theoretical Machine Learning and Quantile Regression'', respectively, in Table 1 of  \cite{ji2016coauthorship}.
When $K= 5, 6,$ and $7$, we can analogously obtain their corresponding communities by combining similar features from $K=8$.
 As discussed in  \cite{ji2016coauthorship}, such decompositions
can not only identify the hot areas and primary authors in research communities, but also enhance  understanding of the research habits and
trends of scientific researchers.
Consequently, our proposed test is applicable for detecting community
structures in this sparse network data, which provides more sensible options for practitioners to select the number of communities.

In sum,  our proposed test
and \cite{hu2021using}'s test lead to the same result in the dense network example, and we also provide more interpretations.
In the sparse network example, our  test successfully detects a reasonable
number of network communities, whereas \cite{hu2021using}'s test rejects all possible hypotheses and fails to yield any conclusion.
Consequently, our proposed method is  applicable for both data sets.

\section{Concluding remarks}

In this paper, we combine \cite{hu2021using}'s approach  with a sampling process
to develop a novel test statistic to test the SBM for both dense and sparse networks.
We demonstrate theoretically that the proposed test statistic
converges to the Type-I extreme value distribution under the null hypothesis,
regardless of the network structure. In addition, we have established
 the asymptotic power properties and introduced  the bootstrap-corrected method and  an augmented test statistic.
Moreover, simulation studies support our theoretical findings and two empirical examples illustrate the usefulness of the proposed method.

It is  of interest to note that our proposed method can be modified to accommodate the
 degree-corrected stochastic block model (DCSBM; \citealt{karrer2011stochastic}). Compared with the SBM,
 the DCSBM  includes a set of node degree parameters $\omega=(\omega_i)_{1\le i\le n}$ for measuring the degree variation of nodes and assumes that $P(A_{ij}=1|Q,g,\omega)=\omega_i\omega_jQ_{g(i)g(j)}$. The log-likelihood function under the DCSBM is,
$\ell(A|Q,g,\omega)=\sum_{i=1}^n d_i\log\omega_i+2^{-1}\sum_{u,v=1}^K(m_{uv}\log Q_{uv}-n_{uv}\log Q_{uv})$, where
$m_{uv}$ and $n_{uv}$ are defined in Subsection 2.1, and $d_i=\sum_{j=1}^n A_{ij}$.
The maximum likelihood estimators $\hat Q^{g_0}$ and $\hat\omega$ can then be obtained by maximizing $\ell(A|Q,g_0,\omega)$ for the given
 hypothetical values of $K_0$ and $g_0$. Accordingly, our proposed method can be applied to the DCSBM
by first replacing  $\hat Q^{g_0}_{g_0(i)g_0(j)}$ with $\hat\omega_i\hat\omega_j\hat Q^{g_0}_{g_0(i)g_0(j)}$ in (\ref{eq:Zhang.hatrho}), and then obtaining statistics analogous to those in
(\ref{eq:hat.gamma}) and (\ref{eq:F0}).
The asymptotic null distribution and the power of the resulting test statistic under the DCSBM can be demonstrated theoretically.
Since the theoretical properties and numerical studies under the DCSBM are almost the same as those of the SBM, we do not present them in the paper.

\section*{Rerefences}

\begin{description}
	\newcommand{\enquote}[1]{``#1''}
	\expandafter\ifx\csname natexlab\endcsname\relax\def\natexlab#1{#1}\fi

	\bibitem[{Agarwal et al.(2013)}]{agarwal2013traffic}
	Agarwal, S., Kodialam, M., and Lakshman, T. (2013),
	\enquote{Traffic Engineering in Software Defined Networks,}
	\textit{2013 Proceedings IEEE INFOCOM}, 2211--2219.

	\bibitem[{Amini et al.(2013)}]{amini2013pseudo}
Amini, A. A., Chen, A., Bickel, P. J., and Levina, E. (2013),
\enquote{Pseudo-Likelihood Methods for Community Detection in Large Sparse Networks,}
\textit{The Annals of Statistics}, 41, 2097--2122.

\bibitem[{Andrikopoulos et~al.(2016)}]{andrikopoulos2016four}
Andrikopoulos, A., Samitas, A., and Kostaris, K. (2016),
\enquote{Four Decades of the Journal of Econometrics: Coauthorship Patterns and Networks,}
\textit{Journal of Econometrics}, 195(1), 23--32.

\bibitem[{Bickel and Chen(2009)}]{bickel2009nonparametric}
Bickel, P. J., and Chen, A. (2009),
\enquote{A Nonparametric View of Network Models and Newman--Girvan and Other Modularities,}
\textit{Proceedings of the National Academy of Sciences}, 106, 21068--21073.

\bibitem[{Bickel and Sarkar(2016)}]{bickel2016hypothesis}
Bickel, P. J., and Sarkar, P. (2016),
\enquote{Hypothesis Testing for Automated Community Detection in Networks,}
\textit{Journal of the Royal Statistical Society Series B: Statistical Methodology}, 78, 253--273.

\bibitem[{Bramoull{\'e} et~al.(2009)}]{bramoulle2009identification}
	Bramoull{\'e}, Y., Djebbari, H., and Fortin, B. (2009),
	\enquote{Identification of Peer Effects through Social Networks,}
	\textit{Journal of Econometrics}, 150(1), 41--55.

\bibitem[{Chatterjee(2015)}]{chatterjee2015matrix}
Chatterjee, S. (2015),
\enquote{Matrix Estimation by Universal Singular Value Thresholding,}
\textit{The Annals of Statistics}, 43, 177--214.

\bibitem[{Chen and Lei(2018)}]{chen2018network}
Chen, K., and Lei, J. (2018),
\enquote{Network Cross-Validation for Determining the Number of Communities in Network Data,}
\textit{Journal of the American Statistical Association}, 113, 241--251.

\bibitem[{Chin et al.(2015)}]{chin2015stochastic}
Chin, P., Rao, A., and Vu, V. (2015),
\enquote{Stochastic Block Model and Community Detection in Sparse Graphs: A Spectral Algorithm with Optimal Rate of Recovery,}
\textit{Conference on Learning Theory}, 391--423.

\bibitem[{Choi et al.(2012)}]{choi2012stochastic}
Choi, D. S., Wolfe, P. J., and Airoldi, E. M. (2012),
\enquote{Stochastic Blockmodels with a Growing Number of Classes,}
\textit{Biometrika}, 99, 273--284.

\bibitem[{Daudin et al.(2008)}]{daudin2008mixture}
Daudin, J. J., Picard, F., and Robin, S. (2008),
\enquote{A Mixture Model for Random Graphs,}
\textit{Statistics and Computing}, 18, 173--183.

\bibitem[{Di Maggio et al.(2019)}]{di2019relevance}
Di Maggio, M., Franzoni, F., Kermani, A., and Sommavilla, C. (2019),
\enquote{The Relevance of Broker Networks for Information Diffusion in the Stock Market,}
\textit{Journal of Financial Economics}, 134, 419--446.

\bibitem[{Fu et al.(2025)}]{Fu2025twosample}
Fu, K., Hu, J., Keita, S., and Liu, H. (2025),
\enquote{Two-Sample Test for Stochastic Block Models Via Maximum Entry-Wise Deviation,}
\textit{Journal of the Korean Statistical Society}, 18, 299--313.

\bibitem[{Gao et al.(2017)}]{gao2017achieving}
Gao, C., Ma, Z., Zhang, A. Y., and Zhou, H. H. (2017),
\enquote{Achieving Optimal Misclassification Proportion in Stochastic Block Models,}
\textit{The Journal of Machine Learning Research}, 18, 1980--2024.

\bibitem[{Gao et al.(2018)}]{gao2018community}
Gao, C., Ma, Z., Zhang, A. Y., and Zhou, H. H. (2018),
\enquote{Community Detection in Degree-Corrected Block Models,}
\textit{The Annals of Statistics}, 46, 2153--2185.

\bibitem[{H\"ardle et~al.(2016)}]{hardle2016tenet}
H\"ardle, W. K., Wang, W., and Yu, L. (2016),
\enquote{TENET: Tail-Event Driven Network Risk,}
\textit{Journal of Econometrics}, 192(2), 499--513.

\bibitem[{Han et al.(2023)}]{han2023universal}
Han, X., Yang, Q., and Fan, Y. (2023),
\enquote{Universal Rank Inference via Residual Subsampling with Application to Large Networks,}
\textit{The Annals of Statistics}, 51, 1109--1133.

\bibitem[{Holland et al.(1983)}]{holland1983stochastic}
Holland, P. W., Laskey, K. B., and Leinhardt, S. (1983),
\enquote{Stochastic Blockmodels: First Steps,}
\textit{Social Networks}, 5, 109--137.

\bibitem[{Hu et al.(2021)}]{hu2021using}
Hu, J., Zhang, J., Qin, H., Yan, T., and Zhu, J. (2021),
\enquote{Using Maximum Entry-Wise Deviation to Test the Goodness of Fit for Stochastic Block Models,}
\textit{Journal of the American Statistical Association}, 116, 1373--1382.

\bibitem[{Ji and Jin(2016)}]{ji2016coauthorship}
Ji, P., and Jin, J. (2016),
\enquote{Coauthorship and Citation Networks for Statisticians,}
\textit{The Annals of Applied Statistics}, 10, 1779--1812.

\bibitem[{Jin(2015)}]{jin2015fast}
Jin, J. (2015),
\enquote{Fast Community Detection by SCORE,}
\textit{The Annals of Statistics}, 43, 57--89.

\bibitem[{Jin et~al.(2024)}]{jin2024mixed}
Jin, J., Ke, Z. T., and Luo, S. (2024),
\enquote{Mixed Membership Estimation for Social Networks,}
\textit{Journal of Econometrics}, 239(2), 105369.

\bibitem[{Jing et al.(2022)}]{jing2022community}
Jing, B. Y., Li, T., Ying, N., and Yu, X. (2022),
\enquote{Community Detection in Sparse Networks Using the Symmetrized Laplacian Inverse Matrix (Slim),}
\textit{Statistica Sinica}, 32.

\bibitem[{Jochmans(2024)}]{jochmans2024nonparametric}
Jochmans, K. (2024),
\enquote{Nonparametric Identification and Estimation of Stochastic Block Models from Many Small Networks,}
\textit{Journal of Econometrics}, 242(2), 105805.

\bibitem[{Karrer and Newman(2011)}]{karrer2011stochastic}
Karrer, B., and Newman, M. E. (2011),
\enquote{Stochastic Blockmodels and Community Structure in Networks,}
\textit{Physical Review E}, 83, 016107.

\bibitem[{Karwa et al.(2016)}]{karwa2016exact}
Karwa, V., Pati, D., Petrovic, S., Solus, L., Alexeev, N., Raic, M., Wilburne, D., Williams, R., and Yan, B. (2016),
\enquote{Exact Tests for Stochastic Block Models,}
\textit{arXiv preprint arXiv:1612.06040}.

\bibitem[{Katona et al.(2011)}]{katona2011network}
Katona, Z., Zubcsek, P. P., and Sarvary, M. (2011),
\enquote{Network Effects and Personal Influences: The Diffusion of an Online Social Network,}
\textit{Journal of Marketing Research}, 48, 425--443.


\bibitem[{Lee and Wilkinson(2019)}]{lee2019review}
Lee, C., and Wilkinson, D. J. (2019),
\enquote{A Review of Stochastic Block Models and Extensions for Graph Clustering,}
\textit{Applied Network Science}, 4, 1--50.

\bibitem[{Lei(2016)}]{lei2016goodness}
Lei, J. (2016),
\enquote{A Goodness-of-Fit Test for Stochastic Block Models,}
\textit{The Annals of Statistics}, 44, 401--424.

\bibitem[{Lei and Rinaldo(2015)}]{lei2015consistency}
Lei, J., and Rinaldo, A. (2015),
\enquote{Consistency of Spectral Clustering in Stochastic Block Models,}
\textit{The Annals of Statistics}, 43, 215--237.

\bibitem[{Lei and Zhu(2017)}]{lei2017generic}
Lei, J., and Zhu, L. (2017),
\enquote{Generic Sample Splitting for Refined Community Recovery in Degree Corrected Stochastic Block Models,}
\textit{Statistica Sinica}, 1639--1659.

\bibitem[{Li et al.(2020)}]{li2020network}
Li, T., Levina, E., and Zhu, J. (2020),
\enquote{Network Cross-Validation by Edge Sampling,}
\textit{Biometrika}, 107, 257--276.

\bibitem[{Mossel et al.(2015)}]{mossel2015reconstruction}
Mossel, E., Neeman, J., and Sly, A. (2015),
\enquote{Reconstruction and Estimation in the Planted Partition Model,}
\textit{Probability Theory and Related Fields}, 162, 431--461.

\bibitem[{Mossel et al.(2018)}]{mossel2018proof}
Mossel, E., Neeman, J., and Sly, A. (2018),
\enquote{A Proof of the Block Model Threshold Conjecture,}
\textit{Combinatorica}, 38, 665--708.

\bibitem[{Musumeci et al.(2018)}]{musumeci2018overview}
Musumeci, F., Rottondi, C., Nag, A., Macaluso, I., Zibar, D., Ruffini, M., and Tornatore, M. (2018),
\enquote{An Overview on Application of Machine Learning Techniques in Optical Networks,}
\textit{IEEE Communications Surveys $\&$ Tutorials}, 21, 1383--1408.

\bibitem[{Newman(2006)}]{newman2006modularity}
Newman, M. E. (2006),
\enquote{Modularity and Community Structure in Networks,}
\textit{Proceedings of the National Academy of Sciences}, 103, 8577--8582.

\bibitem[{Nowicki and Snijders(2001)}]{nowicki2001estimation}
Nowicki, K., and Snijders, T. A. B. (2001),
\enquote{Estimation and Prediction for Stochastic Blockstructures,}
\textit{Journal of the American Statistical Association}, 96, 1077--1087.

\bibitem[{Rohe et al.(2011)}]{rohe2011spectral}
Rohe, K., Chatterjee, S., and Yu, B. (2011),
\enquote{Spectral Clustering and the High-Dimensional Stochastic Blockmodel,}
\textit{The Annals of Statistics}, 39, 1878--1915.

\bibitem[{Saldana et al.(2017)}]{saldana2017many}
Saldana, D. F., Yu, Y., and Feng, Y. (2017),
\enquote{How Many Communities Are There?,}
\textit{Journal of Computational and Graphical Statistics}, 26, 171--181.

\bibitem[{Sarkar and Bickel(2015)}]{sarkar2015role}
Sarkar, P., and Bickel, P. J. (2015),
\enquote{Role of Normalization in Spectral Clustering for Stochastic Blockmodels,}
\textit{The Annals of Statistics}, 43, 962--990.

\bibitem[{Snijders and Nowicki(1997)}]{snijders1997estimation}
Snijders, T. A., and Nowicki, K. (1997),
\enquote{Estimation and Prediction for Stochastic Blockmodels for Graphs with Latent Block Structure,}
\textit{Journal of Classification}, 14, 75--100.



\bibitem[{Wang and Bickel(2017)}]{wang2017likelihood}
Wang, Y. R., and Bickel, P. J. (2017),
\enquote{Likelihood-Based Model Selection for Stochastic Block Models,}
\textit{The Annals of Statistics}, 45, 500--528.

\bibitem[{Westveld and Hoff(2011)}]{westveld2011mixed}
Westveld, A. H., and Hoff, P. D. (2011),
\enquote{A Mixed Effects Model for Longitudinal Relational and Network Data, with Applications to International Trade and Conflict,}
\textit{The Annals of Applied Statistics}, 5, 843--872.


\bibitem[{Wu et al.(2022)}]{wu2022inward}
Wu, Y., Lan, W., Zou, T., and Tsai, C. L. (2022),
\enquote{Inward and Outward Network Influence Analysis,}
\textit{Journal of Business $\&$ Economic Statistic}, 40, 1617--1628.

\bibitem[{Wu et al.(2024a)}]{wu2024atwosample}
Wu, Q., and Hu, J. (2024a),
\enquote{Two-Sample Test of Stochastic Block Models,}
\textit{Computational Statistics $\&$ Data Analysis}, 192, 107903.

\bibitem[{Wu et al.(2024)}]{wu2024twosample}
Wu, Q., and Hu, J. (2024),
\enquote{Two-Sample Test of Stochastic Block Models Via the Maximum Sampling Entry-Wise Deviation,}
\textit{Journal of the Korean Statistical Society}, 53, 617--636.

\bibitem[{Zhang and Zhou(2016)}]{zhang2016minimax}
Zhang, A. Y., and Zhou, H. H. (2016),
\enquote{Minimax Rates of Community Detection in Stochastic Block Models,}
\textit{The Annals of Statistics}, 44, 2522--2280.

\bibitem[{Zhang and Amini(2023)}]{zhang2020adjusted}
Zhang, L., and Amini, A. A. (2023),
\enquote{Adjusted Chi-Square Test for Degree-Corrected Block Models,}
\textit{The Annals of Statistics}, 51, 2366--2385.

\bibitem[{Zhao et al.(2012)}]{zhao2012consistency}
Zhao, Y., Levina, E., and Zhu, J. (2012),
\enquote{Consistency of Community Detection in Networks Under Degree-Corrected Stochastic Block Models,}
\textit{The Annals of Statistics}, 40, 2266--2292.

\end{description}

%

\end{document}